\documentclass[lettersize,journal]{IEEEtran}
\usepackage{amsmath,amssymb,amsfonts}
\usepackage{amsthm}
\newtheorem{theorem}{Theorem}
\usepackage{algorithmic}
\usepackage{algorithm}
\usepackage{array} 
\usepackage{subfigure}
\usepackage{textcomp}
\usepackage{stfloats}
\usepackage{url}
\usepackage{verbatim}
\usepackage{graphicx}
\usepackage{cite}
\usepackage{cite}
\usepackage{graphicx}
\usepackage{times}
\usepackage{latexsym}

\usepackage[mathscr]{eucal}
\usepackage{array}
\usepackage{fancyhdr}
\usepackage{amsmath,amssymb}
\usepackage{bm} 
\usepackage{subfigure}
\usepackage{citesort}
\usepackage{multirow}

\ifCLASSINFOpdf
\else
\fi

\usepackage{xcolor}

\makeatother

\usepackage{tabularx}
\usepackage{float}
\usepackage{xcolor}

\usepackage{caption}
\usepackage{multirow}
\usepackage{diagbox}
\usepackage{makecell}
\usepackage{utfsym}
\usepackage{microtype}

\begin{document}

\title{Robust Energy-Efficient Sleep-Mode Strategy for Multi-RIS-Aided Cell-Free Massive MIMO
}

\author{
	\IEEEauthorblockN{Hongyi Luo,~\IEEEmembership{Student Member,~IEEE}, Wenyu Song,~\IEEEmembership{Student Member,~IEEE}, Daniel K. C. So,~\IEEEmembership{Senior Member,~IEEE}, Zahra Mobini,~\IEEEmembership{Senior Member,~IEEE} and Zhiguo Ding,~\IEEEmembership{Fellow,~IEEE}
    }


    \thanks{This work was supported in part by the University of Manchester - China Scholarship Council joint scholarship under Grant 202406090037. An earlier version of this paper was presented in part at the IEEE Global Communications Conference (GLOBECOM) 2025~\cite{Luosleepmode}.}
    \thanks{Hongyi Luo, Wenyu Song, Daniel K. C. So, Zahra Mobini and Zhiguo Ding are with the Department of Electrical \& Electronics Engineering, The University of Manchester, Manchester, M13 9PL, UK, and Zhiguo Ding is also with Khalifa University, Abu Dhabi, UAE. (e-mail: \{hongyi.luo, wenyu.song, d.so, zahra.mobini, zhiguo.ding\}@manchester.ac.uk).}
}

\maketitle

\begin{abstract}
	With the explosive growth of data traffic and the ubiquitous connectivity of wireless devices, the energy demands of wireless networks have inevitably escalated. 
    Reconfigurable intelligent surface (RIS) has emerged as a promising solution for 6G networks due to its energy efficiency (EE) and low cost, while cell-free massive multiple-input multiple-output (CF-mMIMO) was proposed as an innovative network architecture without fixed cell boundaries to enhance these measures even further.
    However, existing studies often assume consistently high traffic loads, neglecting the dynamic nature of user demand. This can result in underutilized access points (APs) and unnecessary energy expenditure during low-demand periods.
    To tackle the challenge of EE in CF-mMIMO systems during low load periods, this paper proposes a novel energy-efficient transmission scheme that jointly coordinates active APs and multiple passive RISs. Specifically, a dynamic AP sleep-mode strategy is designed, where certain APs are selectively deactivated while nearby RISs assist in maintaining coverage.
    We formulate the EE maximization objective as a fractional programming problem and adopt the Dinkelbach method in conjunction with alternating optimization (AO) to iteratively solve the three coupled subproblems: (i) AP selection via a hybrid branch-and-bound (BnB) and greedy algorithm, (ii) transmit power optimization using a sequential convex approximation (SCA) method, initialized by a heuristic zero-forcing strategy, and (iii) RIS phase shift optimization using gradient projection.
    Simulation results show that the proposed scheme achieves significantly higher EE than existing methods in both low and moderate user scenarios.
\end{abstract}

\IEEEpeerreviewmaketitle

\section{Introduction}

Next-generation wireless networks must deliver ultra-high data rate to meet exploding data traffic demands while drastically reducing energy consumption for sustainability~\cite{WCM2025}.
Cell-free massive multiple-input multiple-output (CF-mMIMO) has emerged as a promising network architecture for future sixth-generation (6G) systems, owing to its user-centric transmission paradigm and the lack of strict cell boundaries~\cite{CFmMIMOsurvey}.
CF-mMIMO can significantly improve system capacity, spectral efficiency (SE) and spatial coverage by deploying a large number of access points (APs) to provide enhanced services to user equipments (UEs). In addition, through optimized signal processing techniques such as coordinated beamforming and joint signal processing, CF-mMIMO can also improve energy efficiency (EE) and effectively mitigate interference, thereby improving quality of service (QoS)\cite{PowerprecodingMaxMin,bashar2019energy}.
However, the advantages of this distributed architecture are also accompanied by its inherent challenges. For instance, system performance can be compromised by unfavorable scattering environments or significant signal attenuation resulting from substantial physical obstructions \cite{van2021reconfigurable}. More critically, the deployment of a large number of APs inevitably leads to huge energy consumption. As noted in \cite{verma2020toward}, with the increase in the number of APs and antennas, the associated transceiver energy consumption of CF-mMIMO networks surges, posing significant challenges to fulfilling the requirements of green communication. Indeed, each AP has a static power consumption even at low load. The accumulated static power from a large number of APs can offset the benefits of high SE, thus degrading the overall EE. This makes CF-mMIMO a ``double-edged sword'': while it can satisfy the skyrocketing demands for higher capacity and unwavering reliability, this performance comes at the cost of substantial energy consumption and consequently, poor overall EE. Therefore, EE research on CF-mMIMO must focus on how to intelligently manage the activity status of many APs to reduce their overall energy consumption \cite{tizikara2025socially}. Hence, improving the EE of CF-mMIMO systems has become a critical task~\cite{chen2023energy}.

Recently, reconfigurable intelligent surface (RIS) has emerged as a promising 6G technology that enhances wireless propagation by passively reflecting signals with tunable phase shifts without requiring power-consuming amplifiers~\cite{shi2024ris}.
By incorporating RIS into wireless networks, one can significantly enhance scattering paths for APs, improve the overall channel conditions, and assist signal transmission~\cite{huang2019reconfigurable}. As a result, higher data rates and improved EE can be achieved~\cite{zhang2021joint,liu2023height,ren2023energy,yaswanth2023energy,zhang2021cell}.
The first to introduce RIS in CF networks is~\cite{zhang2021joint}, where the authors developed a joint precoding framework and demonstrated significant improvement on the overall network capacity compared to conventional CF systems.
Liu et al.~\cite{liu2023height} proposed using RIS to improve the EE by reflecting and strengthening the signal to the desired UE. Ren et al.~\cite{ren2023energy} adjusted the phase offset of the RIS reflection unit to enhance the passive beamforming of the signal, thereby significantly improving the power transfer efficiency and reducing the total system energy consumption. Similarly, the combination of RIS with simultaneous wireless information and power transfer (SWIPT) was proposed in \cite{yaswanth2023energy}, which optimized the active beamforming matrix and the passive beamforming matrix of RIS to achieve significant energy-savings. The study in \cite{zhang2021cell} proposed a hybrid beamforming scheme to maximize the EE of an RIS-aided CF-mMIMO system.

However, traffic demands in real-world systems are time-varying; networks often operate under partial load or even no load during off-peak hours, which can lead to significant energy wastage if all APs remain active~\cite{lopez2022survey}. A well-established solution to this problem is the selective deactivation of redundant APs during low-traffic periods, which can substantially reduce energy consumption~\cite{amine2022energy}. This strategy, commonly known as the  base station (BS)/AP sleep mode, is cost-effective because it can be implemented on existing network infrastructure with minimal adjustments, avoiding major hardware modifications and enabling considerable energy and cost savings~\cite{wang2024base,wu2020power,zhang2018operation,ooi2024joint}.
The research community has explored this area from various angles.
For CF-mMIMO systems, recent studies have proposed sophisticated greedy algorithms to intelligently switch off transmit/receive points (TRPs) based on their contribution to overall EE~\cite{sleepmodepowermodel}, while~\cite{ooi2024joint} investigated the joint optimization of sleep mode, power control, and beamforming, employing surrogate machine learning models to handle the problem’s complexity.
Broader research into cellular networks has also contributed to sleep-mode strategies. For example, traffic prediction has been used to proactively deactivate BSs~\cite{zhu2021joint}, and foundational work has examined the design and analysis of multi-level sleep-mode policies, particularly for small-cell deployments~\cite{liu2015small}. The study in~\cite{lin2021data} considered tidal traffic variations and proposed a bidirectional long short-term memory (BLSTM) network to predict future traffic for BS on/off switching. Similarly, \cite{piovesan2021joint} analyzed the effect of traffic fluctuations on sleep-mode strategies and incorporated renewable energy to determine BS activation decisions. The work in~\cite{kim2018traffic} investigated collaborative state management across multiple BSs, exploiting the spatio-temporal characteristics of traffic to design a joint state management and clustering algorithm based on the arrival traffic queue.  Despite these advances, most existing BS/AP sleep-mode strategies overlook the potential of RIS to mitigate coverage degradation caused by deactivated APs. Thanks to their ultra-low power consumption and reconfigurable reflection capabilities, RISs can sustain coverage for UE in areas previously served by the sleeping APs, thereby maintaining network performance while further improving EE.

In this paper, we investigate a novel energy-efficient CF-mMIMO framework aided by multiple RISs. We propose an energy-efficient sleep-mode strategy for this hybrid network, where a subset of APs dynamically enter into sleep-mode to reduce energy consumption, and the resulting coverage degradation is compensated by passive RIS reflection. To tackle the resultant mixed-integer non-convex optimization problem of jointly selecting APs, allocating transmit powers, and configuring RIS phases, we propose an efficient algorithmic framework based on alternating optimization (AO) and Dinkelbach’s method. The main contributions of this work are summarized as follows:

\begin{itemize}
	\item  We propose an energy-efficient transmission strategy for multi-RIS-aided CF-mMIMO systems, in which the operation of active APs and passive RISs is jointly optimized. By dynamically deactivating underutilized APs (i.e., putting these APs to sleep) and leveraging RISs to sustain coverage, the proposed network can adapt to traffic load variations while significantly reducing energy consumption.
	
	\item We formulate the EE maximization problem as a non-convex fractional program, incorporating the effects of imperfect CSI to reflect practical system limitations. We propose a solution framework based on the Dinkelbach method and AO, which decomposes the original problem into three tightly coupled subproblems solved iteratively: (i) AP selection using branch-and-bound (BnB) and greedy algorithm, (ii) transmit power allocation using a sequential convex approximation (SCA) method, and (iii) RIS phase optimization through a gradient projection (GP) method with numerical differentiation. 

    \item To further reduce the algorithmic complexity in large-scale networks, we design efficient heuristics based on greedy AP selection and a bio-inspired whale optimization algorithm (WOA) for RIS configuration, which achieve near-optimal performance with lower computational cost.

	\item Comprehensive simulations demonstrate that the proposed scheme consistently outperforms conventional CF-mMIMO and other benchmarks in terms of EE.
\end{itemize}

The rest of the paper is organized as follows. Section~\ref{sec2} details the system model, including the channel and power consumption models. Section~\ref{problemformulation} formulates the EE maximization problem. Section~\ref{OptimizationFramework} elaborates on the proposed scheme and the corresponding algorithms to solve the problem. Section~\ref{sectionlowcomplexity} details the low-complexity algorithms. Section~\ref{sec5} presents the simulation results. Finally, Section~\ref{sec6} concludes this paper.

\section{System Model}\label{sec2}

\begin{figure}[!t]
    \centering
	\includegraphics[width=0.9\columnwidth]{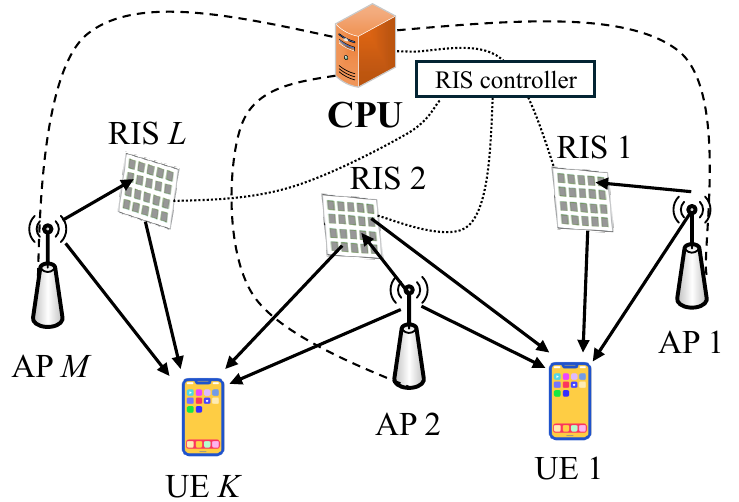}
	\caption{System model of CF-mMIMO with multi-RIS.}
	\label{SystemOverviewAll}
\end{figure}

In this paper, we consider a CF-mMIMO system consisting of $M$ single-antenna APs and $K$ single-antenna UEs,
operating in time-division duplex (TDD) mode, as depicted in Fig.~\ref{SystemOverviewAll}.
The central processing unit (CPU) coordinates the APs, handling system optimization, and signal processing through fronthaul (FH) links.
Moreover, the system is aided by $L$ passive RISs, each connected to a shared remote controller managed by the CPU.
The APs communicate with UEs via a direct path, while the RISs provide a supplementary path by reflecting signals from the APs to the UEs. Each RIS, configured as a uniform planar array (UPA) of $N = N_y \times N_z$ elements, adjusts the phases of incoming signals to reconfigure wave propagation dynamically.
To reduce energy consumption, only a subset of APs indexed by $\mathcal{M}_{\mathcal{A}} \subseteq \{1, \dots, M\}$ is activated to meet service requirements, while others are switched to sleep mode.

\subsection{Channel Model} \label{subsec:channel_model}
We consider a quasi-static block fading model, in which all channels remain constant and frequency-flat within each coherence interval, while varying independently across intervals~\cite{channelmodelal2024performance}.
The composite channel $g_{m k}$ from the $m$-th AP to the $k$-th UE includes a direct link and $L$ reflected links via RISs, given by
\begin{equation}
    g_{m k}=h_{\mathrm{au}, m k}+\sum_{l=1}^L \mathbf{h}_{\mathrm{ar}, m l}^T \boldsymbol{\Phi}_l \mathbf{h}_{\mathrm{ru}, l k},
\end{equation}  
where $h_{\mathrm{au},mk}$ denotes the direct channel between AP $m$ and UE $k$. To capture a realistic propagation environment, this channel is modeled with a probabilistic line-of-sight (LoS). The channel exhibits Rician fading when a LoS path exists, and Rayleigh fading under non-line-of-sight (NLoS) conditions. The Rician fading model is given by
\begin{equation}
    h_{\mathrm{au}, m k} \sim \mathcal{CN} \left( \sqrt{\frac{\beta_{\mathrm{au}, m k} \kappa_{\mathrm{au},mk}}{\kappa_{\mathrm{au},mk} + 1}} e^{j \varphi_{\mathrm{au}, mk}^{\text{AoD}}}, \frac{\beta_{\mathrm{au}, m k}}{\kappa_{\mathrm{au},mk} + 1} \right),
\end{equation}  
where $\beta_{\mathrm{au},mk}$ is the large-scale fading, $\kappa_{\mathrm{au},mk}$ is the Rician factor, and $\varphi_{\mathrm{au},mk}^{\mathrm{AoD}}$ is the angle of departure (AoD) at AP $m$.
The detailed probabilistic model used to determine the LoS/NLoS state and the corresponding value of $\kappa_{au,mk}$ is presented in simulation setup in Section~\ref{simulationsetup}.

We model the AP-RIS channel as a pure LoS link, which is a practical assumption, as both APs and RISs are typically deployed at elevated positions to ensure a clear, unobstructed path~\cite{channelmodelal2024performance,kammoun2020asymptotic}. The channel is expressed as
\begin{equation}
    \mathbf{h}_{\mathrm{ar}, m l} = \sqrt{\beta_{\mathrm{ar}, m l}} \mathbf{a}_N(\theta_{\mathrm{ar}, m l}^{\text{AoA}}, \phi_{\mathrm{ar}, m l}^{\text{AoA}}) \in \mathbb{C}^{N\times1},
\end{equation}  
where $\mathbf{a}_N(\theta,\phi)$ is the RIS array response vector, constructed from the horizontal phase shifts related to angle $\theta$ and the vertical phase shifts related to angle $\phi$, and $\beta_{\mathrm{ar},ml}$ denotes the large-scale fading coefficient.
The two effective angles of arrival (AoAs) from AP $m$ to RIS $l$ are given by $\theta^{\mathrm{AoA}}_{\mathrm{ar},ml}$ and $
\phi^{\mathrm{AoA}}_{\mathrm{ar},ml}$ respectively.

The RIS-UE channel is modeled as spatially correlated Rician fading:
\begin{equation}
    \mathbf{h}_{\mathrm{ru}, l k} \sim \mathcal{CN}\left( \sqrt{\frac{\beta_{\mathrm{ru}, l k} \kappa_{\mathrm{ru},lk}}{\kappa_{\mathrm{ru},lk}+1}} \mathbf{a}_N\left( \theta^{\text{AoD}}_{\mathrm{ru},lk}, \phi^{\text{AoD}}_{\mathrm{ru},lk} \right), \mathbf{\Gamma}_{\mathrm{ru},lk} \right),
\end{equation}
where $\beta_{\mathrm{ru},lk}$ and $\kappa_{\mathrm{ru},lk}$ are the large-scale fading and Rician factor, respectively, and $\mathbf{\Gamma}_{\mathrm{ru},lk}$ is the spatial correlation matrix. Each AoD angle $\theta^{\text{AoD}}_{\mathrm{ru},lk}$ or $\phi^{\text{AoD}}_{\mathrm{ru},lk}$ is derived geometrically using the RIS configuration, element spacing $d_R$, and wavelength $\lambda$.
The covariance matrix characterizing the spatial correlation of NLoS components is given by
$\mathbf{\Gamma}_{\mathrm{ru},lk} = \frac{\beta_{\mathrm{ru},lk}}{\kappa_{\mathrm{ru},lk} + 1} \mathbf{R}$,
where $\mathbf{R} \in \mathbb{C}^{N \times N}$ denotes the RIS spatial correlation matrix.

The $N \times N$ RIS phase shift matrix is defined as
\begin{equation}
    \boldsymbol{\Phi}_l = \operatorname{diag}( [e^{j\theta_{l,1}}, \dots, e^{j\theta_{l,N}}] ),
\end{equation}
where each $\theta_{l,n} \in [0, 2\pi)$ denotes the phase shift induced by the $n$-th element of RIS $l$.


Finally, stacking all APs, the aggregate downlink channel to UE $k$ is written as
\begin{equation}
    \mathbf{g}_{k} = \bar{\mathbf{g}}_{k} + \tilde{\mathbf{g}}_{k},
\end{equation}
where $\mathbf{g}_{k} = [g_{1k}, \ldots, g_{Mk}]^{T} \in \mathbb{C}^{M \times 1}$, the terms $\bar{\mathbf{g}}_{k} = \bar{\mathbf{h}}_{\mathrm{au},k} + \sum_{l=1}^{L} \mathbf{H}_{\mathrm{ar},l} \boldsymbol{\Phi}_{l} \bar{\mathbf{h}}_{\mathrm{ru},lk}$ and $\tilde{\mathbf{g}}_{k} = \tilde{\mathbf{h}}_{\mathrm{au},k} + \sum_{l=1}^{L} \mathbf{H}_{\mathrm{ar},l} \boldsymbol{\Phi}_{l} \tilde{\mathbf{h}}_{\mathrm{ru},lk}$ represent the LoS and NLoS components of the AP-RIS-UE cascaded channel, respectively. The components are defined as follows: $\mathbf{H}_{\mathrm{ar},l} = [\mathbf{h}_{\mathrm{ar},1l}, \ldots, \mathbf{h}_{\mathrm{ar},Ml}]^{T} \in \mathbb{C}^{M \times N}$ is the channel matrix from all APs to RIS $l$, $\bar{\mathbf{h}}_{\mathrm{ru},lk}$ and $\tilde{\mathbf{h}}_{\mathrm{ru},lk}$ are the LoS and NLoS components of the $N \times 1$ channel vector from RIS $l$ to UE $k$, and $\bar{\mathbf{h}}_{\mathrm{au},k}$ and $\tilde{\mathbf{h}}_{\mathrm{au},k}$ are the $M \times 1$ vectors formed by stacking the LoS and NLoS components of the direct channels from all APs to UE $k$.

\subsection{Uplink Training and Channel Estimation}\label{ULtrainingchannelestimation}

We adopt a TDD protocol and utilize the direct estimation (DE) method proposed in~\cite{channelmodelal2024performance}.
The aggregated channel vector $\mathbf{g}_k \in \mathbb{C}^{M \times 1}$ from all APs to UE $k$ is statistically modeled as
\begin{equation}
    \mathbf{g}_k \sim \mathcal{CN}(\bar{\mathbf{g}}_k, \mathbf{Q}_k), \label{eq:gk_statistical}
\end{equation}
where the covariance matrix $
    \mathbf{Q}_k = \sum_{l=1}^L \mathbf{H}_{\mathrm{ar},l} \boldsymbol{\Phi}_l \boldsymbol{\Gamma}_{\mathrm{ru},lk} \boldsymbol{\Phi}_l^H \mathbf{H}_{\mathrm{ar},l}^H + \tilde{\boldsymbol{\beta}}_{\mathrm{au},mk} \mathbf{I}_M$, 
    and $\tilde{\boldsymbol{\beta}}_{\mathrm{au},mk} \in \mathbb{R}^M$ is a vector with entries $[\tilde{\boldsymbol{\beta}}_{\mathrm{au},mk}]_m = \frac{\beta_{\mathrm{au},mk}}{\kappa_{\mathrm{au},mk} + 1}$.
    
Under minimum mean square error (MMSE), the channel estimate and estimation error are distributed as
\begin{equation}
    \hat{\mathbf{g}}_k \sim \mathcal{CN}\left( \bar{\mathbf{g}}_k, \tau_t p_\mathrm{u}  \mathbf{Q}_k \boldsymbol{\Psi}_k \mathbf{Q}_k \right),
\end{equation}
\begin{equation}
    \check{\mathbf{g}}_k = \mathbf{g}_k - \hat{\mathbf{g}}_k \sim \mathcal{CN}\left( \mathbf{0}, \mathbf{C}_k \right),
\end{equation}
where $p_\mathrm{u}$ is the uplink pilot power, $\tau_t$ is the training duration, $\boldsymbol{\Psi}_k = \left( \tau_t p_\mathrm{u} \mathbf{Q}_k + \sigma_u^2 \mathbf{I}_M \right)^{-1}$, where $\sigma_u^2$ is the variance of the additive white Gaussian noise (AWGN) at each AP receiver, assumed to be identical across all APs, and $\mathbf{C}_k = \mathbf{Q}_k - \tau_t p_\mathrm{u} \mathbf{Q}_k \boldsymbol{\Psi}_k \mathbf{Q}_k$ is the estimation error covariance matrix.

\subsection{Power Consumption Model} \label{PowerConsumModel}
The total power consumption in the considered multi-RIS-aided CF-mMIMO system consists of the power consumed by APs $P_{\text{AP}}$, the CPU $P_{\text{CPU}}$, the FH links $P_{\text{FH}}$, and the RISs $P_{\text{RIS}}$~\cite{sleepmodepowermodel}. The overall power consumption is formulated as
\begin{equation}
    P_{\text {total }}= P_{\text{AP}} + P_{\text{CPU}} + P_{\text{FH}} + P_{\text{RIS}}.
\end{equation}

The power consumption of the APs consists of a fixed hardware-related power component and a variable component depending on transmission power, given by  
\begin{equation}\label{PAP}
P_{\text{AP}}= \sum_{m=1}^M (1 - \varpi  (1 - \delta_m)) P_{m}^{\text{AP,Fix}} + \sum_{m=1}^M \delta_m P_{\text{m}}^{\text{AP,TX}},
\end{equation}  
where \( P_{m}^{\text{AP,Fix}} \) represents the fixed power consumption of the $m$-th AP, $\varpi$ denotes the power saving factor of dormancy, which indicates that an AP in sleep mode consumes $\varpi$ less fixed power compared to its active mode~\cite{sleepmodepowermodel}. The term \( P_{\text{m}}^{\text{AP,TX}} \) denotes the transmission power of the \( m \)-th AP, and \( \delta_m \) is a binary indicator, which takes the value 1 if the AP is active and 0 otherwise.

The CPU handles signal processing operations, and its power consumption is expressed as  
\begin{equation}
P_{\text{CPU}}= P_{\text{CPU}}^{\text{Fix}} + B \sum_{k = 1}^K R_{k} P_{\text{CPU}}^{\text{Pre}},
\end{equation}  
where \( P_{\text{CPU}}^{\text{Fix}} \) denotes the fixed CPU power (W), \( P_{\text{CPU}}^{\text{Pre}} \) denotes the precoding power per Gbps of data rate (W/Gbps), $B$ denotes the system bandwidth, and $R_k$ represents the SE of the $k$-th UE. 

The power consumption of the FH links consists of both fixed power $P_{m}^{\text{FH,Fix}}$ required to maintain FH connectivity for the \( m \)-th AP and variable power consumption $P_{m}^{\text{FH,var}}$ per bit in the FH transmission:  
\begin{equation}
P_{\text{FH}}= \sum_{m=1}^M P_{m}^{\text{FH,Fix}} + B \left(\sum_{k = 1}^K R_{k}\right) \sum^{M}_{m=1} \delta_m  P_{m}^{\text{FH,var}},
\end{equation}  
where the indicator $\delta_m$, as defined in (\ref{PAP}), determines whether the corresponding AP contributes to the variable FH power consumption.  

The power consumption of RISs is primarily attributed to the operation of the reflecting elements, which require minimal energy compared to active components like APs. Since RISs passively reflect signals without active transmission or reception, their power consumption is modeled as a fixed overhead:  
$P_{\text{RIS}} = \sum_{l=1}^L P_l^{\mathrm{RIS}}$,
where \( P_l^{\mathrm{RIS}} \) represents the static power consumption of the \( l \)-th RIS.




\section{Problem Formulation} \label{problemformulation}
In this paper, we consider zero-forcing (ZF) precoding over the active APs. ZF is widely used in CF-mMIMO systems~\cite{PowerprecodingMaxMin,channelmodelal2024performance}, offering effective inter-user interference suppression while providing a tractable structure for analyzing performance under imperfect CSI. As our focus lies in the joint optimization of AP sleep-mode, RIS configuration, and power allocation, ZF serves as a convenient and analytically suitable precoding scheme.

Let $\boldsymbol{\Delta} = \mathrm{diag}(\delta_1, \ldots, \delta_M)$ be a binary diagonal matrix indicating the activation status of all APs. The received signal at UE $k$ is given by
\begin{equation}
    r_k = \mathbf{g}_k^T \boldsymbol{\Delta} \mathbf{W}_{\mathrm{ZF}} \mathbf{s} + n_k,
\end{equation}
where $\mathbf{s} = [s_1, \dots, s_K]^T$ contains the unit-energy data symbols, $\mathbf{W}_{\mathrm{ZF}} = \hat{\mathbf{G}}^*(\hat{\mathbf{G}}^T \boldsymbol{\Delta} \hat{\mathbf{G}}^*)^{-1} \mathbf{P}$ represents the ZF precoding matrix, $\hat{\mathbf{G}}$ is the channel estimate matrix, $\mathbf{P} = \mathrm{diag}(\sqrt{p_1}, \dots, \sqrt{p_K})$ is the power allocation matrix, and $n_k \sim \mathcal{CN}(0, \sigma_{d,k}^2)$ is the AWGN at UE $k$, where $\sigma_{d,k}^2$ denotes the downlink noise variance for UE $k$.

Under imperfect channel state information (CSI), the channel is decomposed as $\mathbf{g}_k = \hat{\mathbf{g}}_k + \check{\mathbf{g}}_k$, where $\hat{\mathbf{g}}_k$ is the MMSE estimate and $\check{\mathbf{g}}_k$ is the estimation error. The signal model becomes
\begin{equation}
    r_k = (\hat{\mathbf{g}}_k^T + \check{\mathbf{g}}_k^T) \boldsymbol{\Delta} \hat{\mathbf{G}}^* (\hat{\mathbf{G}}^T \boldsymbol{\Delta} \hat{\mathbf{G}}^*)^{-1} \mathbf{P} \mathbf{s} + n_k,
\end{equation}
where $\hat{\mathbf{G}} = [\hat{\mathbf{g}}_1, \ldots, \hat{\mathbf{g}}_K]$. To approximate the optimal power allocation under imperfect CSI, we adopt a heuristic ZF-based strategy inspired by~\cite{PowerprecodingMaxMin}. Specifically, we set the power allocated to UE $k$ as $p_k = \frac{P_{\max}}{\max_m \sum_{i=k}^{K} \eta_{mi}}$,
where $P_{\max}$ denotes the maximum transmit power per AP, $\eta_{mi}$ is the $i$-th element of the vector $\boldsymbol{\eta}_m = \mathrm{diag}\left\{ \mathbb{E} \left[ (\hat{\mathbf{G}}^T \boldsymbol{\Delta} \hat{\mathbf{G}}^*)^{-1} \hat{\mathbf{g}}_m \hat{\mathbf{g}}_m^H (\hat{\mathbf{G}}^T \boldsymbol{\Delta} \hat{\mathbf{G}}^*)^{-1} \right] \right\}$ and $\hat{\mathbf{g}}_m$ is Hermitian transpose of the $m$-th row of the matrix $\hat{\mathbf{G}}$.

\textit{Remark:} Compared with~\cite{PowerprecodingMaxMin}, our formulation introduces the AP selection mask $\boldsymbol{\Delta}$ to reflect the effect of sleep-mode operation in the precoding and power allocation design. This discussion is limited to the preserved dimensions/subspace (i.e., the direction where the diagonal elements of $\boldsymbol{\Delta}$ are equal to 1), while the masked components are disregarded.


\begin{theorem}
\label{thm:achievable_rate}
For a given UE $k$, under the ZF precoding scheme with imperfect CSI, the instantaneous downlink achievable rate is given by $R_k = \log_2(1 + \text{SINR}_k)$, with the signal-to-interference-plus-noise ratio (SINR) expressed as:
\begin{equation}
    \mathrm{SINR}_k = \frac{p_k}{\sum_{i=1}^{K} p_i \gamma_{k,i} + \sigma_{d,k}^2},
    \label{eq:sinr_theorem}
\end{equation}
where the term $\sum_{i=1}^{K} p_i \gamma_{k,i}$ represents the interference power caused by channel estimation errors, with the error-induced interference term $\gamma_{k,i}$ as the $i$-th element of the vector given by
\begin{equation}
\begin{aligned}
    \boldsymbol{\gamma}_k = \mathrm{diag} \Big\{ \mathbb{E} \Big[  
    & \left( (\hat{\mathbf{G}}^T \boldsymbol{\Delta} \hat{\mathbf{G}}^*)^{-1} \right)^H 
    \hat{\mathbf{G}}^T \boldsymbol{\Delta}  \mathbf{C}_k \boldsymbol{\Delta} \\
    & \hat{\mathbf{G}}^* (\hat{\mathbf{G}}^T \boldsymbol{\Delta} \hat{\mathbf{G}}^*)^{-1} 
    \Big] \Big\}.
\end{aligned}
\label{eq:gamma_theorem}
\end{equation}
\end{theorem}

\begin{IEEEproof}
The detailed proof is provided in Appendix \ref{appendix:proof_of_rate_theorem}.
\end{IEEEproof}

\begin{figure*}[ht]
\begin{equation} \label{Objectiveinitial}
    EE = \frac{B \sum_{k=1}^{K} R_k}{P_{\mathrm{total}}} = \frac{B \sum_{k=1}^K \log_2\left(1+ \frac{p_k}{ \sum_{i=1}^K p_i \gamma_{k,i} + \sigma_{d,k}^2}\right)}{\sum_{m=1}^M \delta_m\left( 
 \varpi P_{m}^{\text{AP,Fix}} +  P_{m}^{\text{AP,TX}} + B \left(\sum_{k = 1}^K R_{k}\right) P_{m}^{\text{FH,var}} \right) + B \left(\sum_{k = 1}^K R_{k}\right) P_{\text{CPU}}^{\text{Pre}} + P_{\text{Fix}}},
\end{equation}
\hrulefill
\vspace*{-2mm}
\end{figure*}
The system-wide EE is defined as (\ref{Objectiveinitial}).
For notational simplicity, we denote the total fixed power component by $P_{\mathrm{Fix}} = \sum_{m=1}^M (1 - \varpi) P_{m}^{\mathrm{AP, Fix}} + \sum_{m=1}^M P_{m}^{\mathrm{FH,Fix}} + \sum_{l=1}^L P_l^{\mathrm{RIS}} + P_{\mathrm{CPU}}^{\mathrm{Fix}}$.
We thus formulate the EE maximization problem as
\begin{align}
\max_{\{\boldsymbol{\Delta}, \boldsymbol{\Phi}, \mathbf{P} \}} & \quad EE \label{Problem2}\\
\mathrm{s.t.} & \quad R_k \geq \xi_k, \quad \forall k, \tag{\ref{Problem2}{a}} \label{Problem2a}\\
& \quad \delta_m \in \left\{0,1\right\}, \quad m = 1, \dots, M, \tag{\ref{Problem2}{b}} \label{Problem2b}\\
& \quad |\phi_{l,n}| = 1, \quad l = 1, \dots, L, n = 1, \dots, N, \tag{\ref{Problem2}{c}} \label{Problem2c}\\
& \quad 0 < p_k \leq P_{\text{max}}^{\text{UE}}, \quad k = 1, \dots, K, \tag{\ref{Problem2}{d}}\label{Problem2d}\\
& \quad P_{m}^{\text{AP,TX}} \leq P_{\text{max}}, \quad \forall m \in \mathcal{M}_{\mathcal{A}}, \tag{\ref{Problem2}{e}}
\label{Problem2e}
\end{align}
where $\{\boldsymbol{\Delta}, \boldsymbol{\Phi}, \mathbf{P} \}$ are the optimization variables representing the AP selection, the RIS phase-shift configuration, and the user power allocation, respectively. Constraint \eqref{Problem2a} is the QoS constraint, ensuring that the data rate for each UE $k$ does not fall below a minimum requirement $\xi_k$. Constraint \eqref{Problem2b} is the binary sleep-mode indicator for each AP.
Constraint \eqref{Problem2c} is the unit-modulus constraint applied to each element $\phi_{l,n}$ of the RIS phase-shift matrices $\boldsymbol{\Phi} = \{\boldsymbol{\Phi}_1, \dots, \boldsymbol{\Phi}_L\}$, reflecting the passive nature of the reflecting elements.
Constraint \eqref{Problem2d} bounds the power parameter $p_k$ for each UE, which not only ensures a positive power allocation but also imposes an upper limit $P_{\text{max}}^{\text{UE}}$ to maintain fairness among UEs and enhance the numerical stability of the optimization algorithm.
Finally, constraint \eqref{Problem2e} represents the maximum transmit power limit $P_{\text{max}}$ per-AP.

\section{Optimization Framework for AP Sleep-Mode, Power Allocation, and RIS Phase Shift}\label{OptimizationFramework}
In this section, we detail the optimization framework to maximize the EE of the multi-RIS-aided CF-mMIMO system. This involves jointly optimizing AP sleep-mode control ($\boldsymbol{\Delta}$), downlink power allocation ($\mathbf{P}$), and RIS phase shifts ($\boldsymbol{\Phi}$).

\subsection{Dinkelbach Method with Alternating Optimization}


The EE maximization problem formulated in (\ref{Problem2}) is a non-convex fractional program. Furthermore, the problem involves mixed-integer variables ($\delta_m \in \{0,1\}$) for AP sleep-mode control, continuous variables for RIS phase shifts, and positive power allocation variables, all coupled in both the numerator and denominator. To address this complexity, we adopt a three-layer optimization framework that combines the Dinkelbach method and AO.

To handle the fractional form of the EE objective function in optimization problem (\ref{Problem2}),
\begin{equation}
    \max_{\{\boldsymbol{\Delta}, \mathbf{P}, \boldsymbol{\Phi}\}} \quad \frac{f(\boldsymbol{\Delta}, \mathbf{P}, \boldsymbol{\Phi})}{g(\boldsymbol{\Delta}, \mathbf{P}, \boldsymbol{\Phi})},
\end{equation}
we transform it into a sequence of parametric subtractive-form problems using the Dinkelbach method as:
\begin{equation}\label{Dinkelbachproblem}
    \max_{\{\boldsymbol{\Delta}, \mathbf{P}, \boldsymbol{\Phi}\}} \quad f(\{\boldsymbol{\Delta}, \mathbf{P}, \boldsymbol{\Phi}\}) - \alpha^{(t)} g(\{\boldsymbol{\Delta}, \mathbf{P}, \boldsymbol{\Phi}\}),
\end{equation}
where $f(\cdot)$ denotes the sum rate (numerator in (\ref{Objectiveinitial}))
and $g(\cdot)$ denotes the total power consumption (denominator in (\ref{Objectiveinitial})).
The parameter $\alpha^{(t)}$ represents the EE value at iteration $t$ and is updated as $\alpha^{(t+1)} = f(\boldsymbol{\Delta}^{(t)}, \mathbf{P}^{(t)}, \boldsymbol{\Phi}^{(t)}) / g(\boldsymbol{\Delta}^{(t)}, \mathbf{P}^{(t)}, \boldsymbol{\Phi}^{(t)})$ after solving problem (\ref{Dinkelbachproblem}) to obtain the optimal variables for that iteration, until convergence.


Even after this transformation, problem (\ref{Dinkelbachproblem}) remains highly non-convex due to the joint optimization over the three sets of variables \(\{\boldsymbol{\Delta}, \mathbf{P}, \boldsymbol{\Phi}\}\). To handle this, we employ an AO strategy within each Dinkelbach iteration. We cyclically optimize one group of variables while keeping the others fixed:
\begin{itemize}
    \item {AP selection:} For fixed RIS phase shifts $\boldsymbol{\Phi}$ and power allocation $\mathbf{P}$, solve a binary optimization problem to determine the optimal AP sleep-mode vector $\boldsymbol{\Delta}$.
    \item {Power allocation:} For fixed AP selection $\boldsymbol{\Delta}$ and RIS phase shifts $\boldsymbol{\Phi}$, allocate transmit power $\mathbf{P}$ among UEs based on SCA method under imperfect CSI.
    \item {RIS phase shift design:} For fixed AP selection $\boldsymbol{\Delta}$ and power allocation $\mathbf{P}$, solve a continuous optimization problem to tune the RIS reflection coefficients $\boldsymbol{\Phi}$ to improve the overall rate.
\end{itemize}
Within each Dinkelbach iteration, the AO algorithm cycles through the subproblems of AP selection, power allocation, and RIS phase optimization. The solution obtained at the end of one full AO cycle is then used as the input for the next cycle. This iterative process is repeated until the improvement in the objective of (\ref{Dinkelbachproblem}) is below a predefined threshold.

\subsection{Subproblem 1: Fix $(\boldsymbol{\Phi}, \mathbf{P})$ and Solve $\boldsymbol{\Delta}$}\label{sectionsubproblem1}

In this subsection, we consider the subproblem of optimizing the AP selection variables $\boldsymbol{\Delta}$, using the phase-shift variables $\boldsymbol{\Phi}$ and power allocation variables $\mathbf{P}$ obtained from the previous AO iteration.
By substituting these fixed $\boldsymbol{\Phi}$ and $\mathbf{P}$ into problem \eqref{Dinkelbachproblem}, the subproblem reduces to optimizing $\boldsymbol{\Delta}$:
\begin{align}
\max_{\boldsymbol{\Delta}} & \quad 
f(\boldsymbol{\Delta}) \;-\; \alpha^{(t)}\,g(\boldsymbol{\Delta}) \label{Problem1_Relabel}\\
\mathrm{s.t.} 
        & \quad R_k(\boldsymbol{\Delta}) \;\geq\; \xi_k, \quad \forall k, \tag{\ref{Problem1_Relabel}{a}} \label{Problem1_Relabela}\\
& \quad \delta_m \in \{0,1\}, \quad m = 1, \dots, M. \tag{\ref{Problem1_Relabel}{b}} \label{Problem1_Relabelb}
\end{align}

Given the binary and combinatorial nature of the subproblem, we adopt the BnB method to obtain the global optimum in moderate-sized scenarios~\cite{luong2017optimal}. BnB explores the solution space by recursively branching on binary variables \(\delta_m\) and pruning inferior branches based on upper bounds, reducing complexity. 
To further alleviate computational complexity in large-scale systems, we propose a simple yet effective greedy algorithm, as detailed in Section~\ref{lowcomplexGreedy}.


\subsection{Subproblem 2: Fix $(\boldsymbol{\Phi}, \boldsymbol{\Delta})$ and Solve $\mathbf{P}$}

Given the AP selection matrix $\boldsymbol{\Delta}^\star$ (solutions from Subproblem 1) and RIS phase shifts $\boldsymbol{\Phi}^\star$ (solutions from Subproblem 3 of the previous AO iteration, or an initial value), this subproblem optimizes the power allocation vector $\mathbf{P}$ by the SCA method.
A good initial point for the SCA iterations can be obtained using the heuristic strategy from~\cite{PowerprecodingMaxMin}.
The objective function is to maximize the Dinkelbach-transformed function:
\begin{equation} \label{subproblem21}
    \begin{aligned}
        \max_{\{\mathbf{P} \}} & \quad f(\mathbf{P}, \boldsymbol{\Delta}^\star, \boldsymbol{\Phi}^\star) - \alpha^{(t)} g(\mathbf{P}, \boldsymbol{\Delta}^\star, \boldsymbol{\Phi}^\star).
    \end{aligned}   
\end{equation}

   

The sum rate term $f({\mathbf{P}})$ is given by:
\begin{equation}
    f({\mathbf{P}}) = B {\sum_{k=1}^K \log_2\left(1+ \frac{p_k}{ \sum^K_{i=1} p_i \gamma_{k,i}(\boldsymbol{\Phi}^\star) + \sigma_{d,k}^2}\right)},
\end{equation}
where $\gamma_{k,i}(\boldsymbol{\Phi}^\star)$ explicitly shows its dependence on the fixed RIS phases $\boldsymbol{\Phi}^\star$,
as defined in \eqref{eq:gamma_theorem}.
The power consumption part $g({\mathbf{P}})$ can be written as:
\begin{equation}\label{gpinitial25}
    \begin{aligned}
        g({\mathbf{P}}) &= \sum_{m \in \mathcal{M}_{\mathcal{A}}} \left( 
 \varpi P_{m}^{\text{AP,Fix}} +  P_{m}^{\text{AP,TX}} \right) \\ &+ f({\mathbf{P}}) \left(\sum_{m \in \mathcal{M}_{\mathcal{A}}} P_{m}^{\text{FH,var}} + P_{\text{CPU}}^{\text{Pre}} \right) + P_{\text{Fix}}^\prime,
    \end{aligned} 
\end{equation}
where ${\mathcal{M}_{\mathcal{A}}}$ is the set of active APs, and $P_{\mathrm{Fix}}^\prime = \sum_{m \in \mathcal{M}_{\mathcal{A}}} (1 - \varpi) P_{m}^{\mathrm{AP, Fix}} + \sum_{m \in \mathcal{M}_{\mathcal{A}}} P_{m}^{\mathrm{FH,Fix}} + \sum_{l=1}^L P_l^{\mathrm{RIS}} + P_{\mathrm{CPU}}^{\mathrm{Fix}}$.
With ZF precoding $\mathbf{W}_{\text{ZF}}=\mathbf{V}\mathbf{P}$, the power transmitted from AP $m$ is $P_{m}^{\text{AP,TX}}=\sum_{k=1}^{K}p_{k}|[\mathbf{v}_{k}]_{m}|^{2}$, where $\mathbf{V} = \hat{\mathbf{G}}^*(\hat{\mathbf{G}}^T \boldsymbol{\Delta}^\star \hat{\mathbf{G}}^*)^{-1}$, $\mathbf{v}_k$ is the $k$-th column of $\mathbf{V}$.
The total transmit power across all active APs is therefore obtained by summing $P_{m}^{\text{AP,TX}}$ over all $m \in \mathcal{M}_{\mathcal{A}}$, which yields the direct relationship $\sum_{m \in \mathcal{M}_{\mathcal{A}}}P_{m}^{\text{AP,TX}} = \sum_{k=1}^{K}c_{k}p_{k}$. The coefficient $c_{k} \triangleq \sum_{m \in \mathcal{M}_{\mathcal{A}}}|[\mathbf{v}_{k}]_{m}|^{2}$ here represents the cumulative beamforming gain for UE $k$ from all active APs. Based on this result, the total power consumption in (\ref{gpinitial25}) can be rewritten as:
\begin{equation}
    \begin{aligned}
 g(\mathbf{P}) &= \sum_{k=1}^K c_k p_k + f(\mathbf{P}) \underbrace{\left(\sum_{m \in \mathcal{M}_{\mathcal{A}}} P_{m}^{\text{FH,var}} + P_{\text{CPU}}^{\text{Pre}} \right)}_{P_{\text{dyn}}} \\ & + \underbrace{\left( \sum_{m \in \mathcal{M}_{\mathcal{A}}} \varpi P_{m}^{\text{AP,Fix}} + P_{\text{Fix}}^\prime \right)}_{P_{\text{stat}}},
    \end{aligned} 
\end{equation}
where $P_{\text{dyn}}$ represents the sum of coefficients for the power consumption components that are directly proportional to the sum rate $f({\mathbf{P}})$, ${P_{\text{stat}}}$ represents the total effective static power consumption of the system.

The objective function to maximize EE by $\mathbf{P}$ becomes
\begin{equation} \label{subproblem22}
    \begin{aligned}
        \max_{\{\mathbf{P} \}} & \quad  f({\mathbf{P}}) \left(1 - \alpha^{(t)} P_{\text{dyn}} \right) -\alpha^{(t)} \left(\sum_{k=1}^K c_k p_k + {P_{\text{stat}}} \right),
    \end{aligned}   
\end{equation}
which can be reformulated as
\begin{align}
\max_{\{p_k\}} \quad & f(\mathbf{P}) (1 - \alpha^{(t)} P_{\text{dyn}}) - \alpha^{(t)} \sum_{k=1}^K c_k p_k \label{subproblem23}\\
\mathrm{s.t.} \quad & 0 < p_k \leq P_\text{max}^\text{UE}, \quad \forall k, \tag{\ref{subproblem23}{a}} \label{subproblem23a}\\
&\sum_{j=1}^K p_j \left|[\mathbf{v}_j]_m\right|^2 \leq P_{\text{max}}, \quad \forall m \in \mathcal{M}_{\mathcal{A}}, \tag{\ref{subproblem23}{b}} \label{subproblem23b}\\
&B \log_2\left(1+ \frac{p_k}{ \sum_{i=1}^K p_i \gamma_{k,i}(\boldsymbol{\Phi}^\star) + \sigma_{d,k}^2}\right) \geq \xi_k, \quad \forall k, \tag{\ref{subproblem23}{c}} \label{subproblem23c}
\end{align}
where the constraints are inherited from the main problem (\ref{Problem2}). Specifically, (\ref{subproblem23}{a}) and (\ref{subproblem23}{b}) represent the per-UE and per-AP power limits, respectively, with maximum values of $P_\text{max}^\text{UE}$ and $P_{\text{max}}$. Constraint (\ref{subproblem23}{c}) is the QoS requirement for each UE.

The power allocation subproblem is non-convex due to the sum rate term $f(\mathbf{P})$ and the individual rate constraints, $R_k(\mathbf{P}) \geq \xi_k$. To address this, we employ the SCA method, which iteratively solves a sequence of convex subproblems that approximate the original problem. The complete iterative procedure is formally outlined in Algorithm \ref{alg:SCA_Power_Allocation}.
Let $\mathbf{P}^{(s)}$ be the power allocation vector at the $s$-th SCA iteration.

\begin{algorithm}[!t]
\caption{SCA-Based Power Allocation Algorithm}
\label{alg:SCA_Power_Allocation}
\begin{algorithmic}[1]
\STATE \textbf{Input:} 
    Fixed AP selection $\boldsymbol{\Delta}^\star$ and RIS phases $\boldsymbol{\Phi}^\star$;
    Current Dinkelbach parameter $\alpha^{(t)}$;
    Initial power vector $\mathbf{P}^{(0)}$;
    Tolerance $\epsilon_{\text{SCA}}$, max iterations $S_{\text{max}}$.

\STATE \textbf{Initialize:} Set SCA iteration counter $s \leftarrow 0$.

\FOR{$s = 0$ to $S_{\text{max}}-1$}
    \STATE For each UE $k$, construct the concave lower-bound rate $\tilde{R}_k^{(s)}(\mathbf{P})$ at the current point $\mathbf{P}^{(s)}$ as defined in (\ref{eq:rate_lower_bound}).
    
    \STATE Solve the convex subproblem (\ref{eq:sca_power_subproblem}) to obtain the updated power vector $\mathbf{P}^{(s+1)}$.
    
    \IF{$\| \mathbf{P}^{(s+1)} - \mathbf{P}^{(s)} \|_2 / \| \mathbf{P}^{(s)} \|_2 < \epsilon_{\text{SCA}}$}
        \STATE \textbf{Break}
    \ENDIF
\ENDFOR

\STATE \textbf{Return} $\mathbf{P}^* \leftarrow \mathbf{P}^{(s+1)}$.
\end{algorithmic}
\end{algorithm}

Our SCA approach is based on the difference of convex functions (DC) representation of the rate function. The rate of UE $k$ can be rewritten as:
\begin{equation}
R_k(\mathbf{P}) = B \left[ \log_2\left(p_k + D_k(\mathbf{P})\right) - \log_2\left(D_k(\mathbf{P})\right) \right],
\end{equation}
where $D_k(\mathbf{P}) = \sum_{i=1}^K p_i \gamma_{k,i}(\boldsymbol{\Phi}^\star) + \sigma_{d,k}^2$ is affine in $\mathbf{P}$. This expression is a difference of two concave functions, which is non-convex.

The standard SCA technique for DC programming is to linearize the subtracted concave term. At iteration $s$, we replace $\log_2(D_k(\mathbf{P}))$ with its first-order Taylor expansion around $\mathbf{P}^{(s)}$. This expansion provides a global upper bound for $\log_2(D_k(\mathbf{P}))$, thus yielding a concave lower bound for the original rate function $R_k(\mathbf{P})$. This lower bound, denoted as $\tilde{R}_k^{(s)}(\mathbf{P})$, is given by:
\begin{equation}
    \begin{aligned} \label{eq:rate_lower_bound}
\tilde{R}_k^{(s)}(\mathbf{P}) = B \bigg[ & \log_2\left(p_k + D_k(\mathbf{P})\right) - \log_2(D_k(\mathbf{P}^{(s)})) \\
& - \sum_{j=1}^K \frac{\gamma_{k,j}(\boldsymbol{\Phi}^\star)}{\ln 2  D_k(\mathbf{P}^{(s)})} (p_j - p_j^{(s)}) \bigg].
    \end{aligned}
\end{equation}

The detailed derivation of this lower bound is provided in Appendix~\ref{Derivation}. The function $\tilde{R}_k^{(s)}(\mathbf{P})$ is concave because it is the sum of a concave function ($\log_2$ of an affine argument) and an affine function. Consequently, the approximated sum rate $\tilde{f}^{(s)}(\mathbf{P}) = \sum_{k=1}^K \tilde{R}_k^{(s)}(\mathbf{P})$ is also concave.

At the $(s+1)$-th SCA iteration, we solve the following convex optimization problem:
\begin{align}
\max_{\{p_k\}} \quad & \tilde{f}^{(s)}(\mathbf{P}) (1 - \alpha^{(t)} P_{\text{dyn}}) - \alpha^{(t)} \sum_{k=1}^K c_k p_k \label{eq:sca_power_subproblem} \\
\mathrm{s.t.} \quad & 0 < p_k \leq P_\text{max}^\text{UE}, \quad \forall k, \tag{\ref{eq:sca_power_subproblem}{a}} \label{eq:sca_power_c1} \\
& \sum_{j=1}^K p_j |[\mathbf{v}_j]_m|^2 \leq P_{\text{max}}, \quad \forall m \in \mathcal{M}_{\mathcal{A}}, \tag{\ref{eq:sca_power_subproblem}{b}} \label{eq:sca_power_c2} \\
& B \tilde{R}_k^{(s)}(\mathbf{P}) \geq \xi_k, \quad \forall k. \tag{\ref{eq:sca_power_subproblem}{c}} \label{eq:sca_power_c3}
\end{align}
This problem maximizes a concave objective function subject to linear constraints, which is a standard convex optimization problem. It can be solved to its global optimum efficiently using off-the-shelf solvers such as Gurobi~\cite{gurobi}. The solution of (\ref{eq:sca_power_subproblem}) provides the updated power vector $\mathbf{P}^{(s+1)}$, and the SCA iterations continue until convergence is achieved.

\subsection{Subproblem 3: Fix $(\boldsymbol{\Delta},\mathbf{P})$ and Solve $\boldsymbol{\Phi}$}
\begin{algorithm}[t]
\caption{Gradient Projection for RIS Phase Optimization}
\label{alg:GP_RIS_Optimization}
\begin{algorithmic}[1]
\STATE \textbf{Input:} 
    Fixed AP selection $\boldsymbol{\Delta}^*$, power allocation $\mathbf{P}^*$; Initial RIS phases $\boldsymbol{\Phi}^{(0)}$;
    Finite-difference perturbation $\epsilon_{\text{FD}}$; Convergence tolerance $\varepsilon$; Max iterations $T_{\max}$.

\STATE \textbf{Initialize:} Set iteration counter $t \leftarrow 0$.

\FOR{$t = 0$ to $T_{\max}-1$}
    \STATE Compute the numerical gradient of the sum rate, $\nabla f(\boldsymbol{\Phi}^{(t)})$, using the finite-difference method in (\ref{eq:finite_diff}).
    
    \STATE Determine the step size $\alpha^{(t)}$ via Armijo line search.
    
    \STATE Update all RIS phases to obtain $\boldsymbol{\Phi}^{(t+1)}$ by applying the projected gradient ascent rule from (\ref{eq:gradient_projection}) to each element.
    
    \IF{$|f(\boldsymbol{\Phi}^{(t+1)}) - f(\boldsymbol{\Phi}^{(t)})|/|f(\boldsymbol{\Phi}^{(t)})| < \varepsilon$}
        \STATE \textbf{Break}
    \ENDIF
\ENDFOR

\STATE \textbf{Return} $\boldsymbol{\Phi}^* \leftarrow \boldsymbol{\Phi}^{(t+1)}$.
\end{algorithmic}
\end{algorithm}
In this subsection, we fix the optimized AP selection matrix $\boldsymbol{\Delta^*}$ and the power allocation vector $\mathbf{P^*}$. The original Dinkelbach-type problem can be reformulated into an equivalent sum rate maximization problem. Subsequently, we employ an iterative algorithm based on GP to efficiently optimize the RIS phase shifts $\boldsymbol{\Phi}$~\cite{GPmethods}.

With $\boldsymbol{\Delta^*}$ and $\mathbf{P^*}$ fixed, problem (\ref{Dinkelbachproblem}) in iteration $t$ is given by:
\begin{equation} \label{RISoptObj}
    \max_{\boldsymbol{\Phi}} \quad f(\boldsymbol{\Phi}) - \alpha^{(t)} g(\boldsymbol{\Phi}).
\end{equation}
Since the power consumption model is dependent on the rate, $g(\boldsymbol{\Phi})$ has a linearly relationship with $f(\boldsymbol{\Phi})$. Due to the nature of Dinkelbach algorithm, it can easily be derived that maximizing (\ref{RISoptObj}) is equivalent to maximizing $f(\boldsymbol{\Phi})$.
Therefore, problem (\ref{RISoptObj}) can be equivalently reformulated as a sum rate maximization problem
\begin{equation} \label{eq:phase_opt}
\begin{aligned}
\max_{\boldsymbol{\Phi}} \quad & f(\boldsymbol{\Phi}) = B\sum_{k=1}^K \log_2\left(1 + \frac{p_k}{\sum_{i=1}^K p_i \gamma_{k,i}(\boldsymbol{\Phi}) + \sigma_{d,k}^2}\right) \\
\mathrm{s.t.} \quad & |\phi_{l,n}| = 1, \quad \forall l \in \{1,\dots,L\}, n \in \{1,\dots,N\}.
\end{aligned}
\end{equation}

The gradient of the objective function with respect to the RIS phase shifts $\phi_{l,n}$ is expressed by applying the chain rule as: \begin{equation}\label{eq:general_gradient} \frac{\partial f(\boldsymbol{\Phi})}{\partial \phi_{l,n}}=\frac{B}{\ln2}\sum_{k=1}^{K}\frac{p_k}{\text{SINR}_k(\boldsymbol{\Phi})}\frac{\partial \text{SINR}_k(\boldsymbol{\Phi})}{\partial \phi_{l,n}}. \end{equation}

However, due to the inherent complexity of the channel estimation error-induced interference term $\gamma_{k,i}(\boldsymbol{\Phi})$, a closed-form expression of $\frac{\partial \text{SINR}_k(\boldsymbol{\Phi})}{\partial \phi_{l,n}}$ gradient is difficult to derive.
Consequently, we employ a numerical finite-difference approximation to compute this gradient: \begin{equation}\label{eq:finite_diff} \frac{\partial f(\boldsymbol{\Phi})}{\partial \theta_{l,n}}\approx\frac{f(\theta_{l,n}+\epsilon)-f(\theta_{l,n}-\epsilon)}{2\epsilon}, \end{equation} where $\epsilon$ denotes a sufficiently small perturbation in the phase shift.

Using the numerically computed gradients, the RIS phase shifts are iteratively updated through gradient ascent followed by a projection onto the feasible set defined by unit-modulus constraints: \begin{equation}\label{eq:gradient_projection} \phi_{l,n}^{(t+1)}=\exp\left(j\arg\left(\phi_{l,n}^{(t)}+\alpha^{(t)}\frac{\partial f(\boldsymbol{\Phi})}{\partial \phi_{l,n}}\right)\right), \end{equation} 
where the step size $\alpha^{(t)}$ is adaptively selected using an Armijo-type line search to ensure stable convergence and monotonic improvement~\cite{armijo1966minimization}.
The iterative framework continues until the incremental improvement in sum rate satisfies: \begin{equation}\label{eq:convergence_criterion} |f(\boldsymbol{\Phi}^{(t+1)})-f(\boldsymbol{\Phi}^{(t)})|/|f(\boldsymbol{\Phi}^{(t)})| < \varepsilon, \end{equation} 
where $\varepsilon$ is a predefined convergence threshold. 

This numerical framework eliminates the need for explicit channel state information in gradient calculations, making it robust to practical implementation uncertainties. The complete algorithm is summarized in Algorithm~\ref{alg:GP_RIS_Optimization}.
The optimized phase shifts $\boldsymbol{\Phi}^*$ obtained here will then be used as the input for Subproblem 1 in the next AO iteration. This process continues in an iterative manner until the objective function in \eqref{Dinkelbachproblem} converges.

\section{Scalable Low-Complexity Heuristic Algorithms}\label{sectionlowcomplexity}
The framework developed in Section~\ref{OptimizationFramework}, which we term the near-optimal solution, establishes a high-performance benchmark by employing rigorous methods like BnB to find the near-optimal solution for certain subproblems. However, this approach may incur significant computational complexity in large-scale scenarios.
In this section, we present the low-complexity solution, which utilizes alternative heuristic algorithms for AP selection and RIS phase shift optimization. The aim is to drastically reduce the computational overhead while maintaining a comparable performance to the near-optimal solution.


\subsection{Greedy-Based AP Selection for Subproblem 1}\label{lowcomplexGreedy}

\begin{algorithm}[!t]
\caption{Greedy Heuristic for AP Selection}
\label{Alg:Greedy}
\begin{algorithmic}[1]
\STATE \textbf{Input:} Fixed phase-shift matrix $\boldsymbol{\Phi}^*$, power allocation matrix $\mathbf{P}^*$; Dinkelbach parameter $\alpha^{(t)}$; QoS constraints $\{\xi_k\}$.

\STATE \textbf{Initialize:} Set of active APs $\mathcal{M}_{\mathcal{A}} \leftarrow \{1, \dots, M\}$.

\WHILE{true}
    \STATE Find candidate set $\mathcal{C} \leftarrow \{ m \in \mathcal{M}_{\mathcal{A}} \mid R_k(\mathcal{M}_{\mathcal{A}} \setminus \{m\}) \geq \xi_k, \forall k \}$.
    
    \IF{$\mathcal{C}$ is empty} \STATE \textbf{Break}; \ENDIF

    \STATE $m^* \leftarrow \arg\max_{m \in \mathcal{C}} J(\mathcal{M}_{\mathcal{A}} \setminus \{m\})$.
    
    \IF{$J(\mathcal{M}_{\mathcal{A}} \setminus \{m^*\}) > J(\mathcal{M}_{\mathcal{A}})$}
        \STATE $\mathcal{M}_{\mathcal{A}} \leftarrow \mathcal{M}_{\mathcal{A}} \setminus \{m^*\}$.
    \ELSE
        \STATE \textbf{Break};
    \ENDIF
\ENDWHILE

\STATE Construct $\boldsymbol{\Delta}^*$ from the final set $\mathcal{M}_{\mathcal{A}}$ (i.e., $[\boldsymbol{\Delta}^*]_{mm}=1$ if $m \in \mathcal{M}_{\mathcal{A}}$, else 0).
\STATE \textbf{Return} $\boldsymbol{\Delta}^*$.
\end{algorithmic}
\end{algorithm}

While the BnB method can find the optimal solution to the AP selection subproblem~\eqref{Problem1_Relabel}, its exponential complexity makes it computationally infeasible for networks with a large number of APs. To address this scalability challenge, we propose a low-complexity greedy turn-off heuristic, as detailed in Algorithm \ref{Alg:Greedy}.
The core idea is to start with all APs active and iteratively deactivate the single AP that provides the most significant improvement in the objective function, until no further gains are possible.


For notational clarity within the algorithm description, we define the subproblem 1 (\ref{Problem1_Relabel}) for a given active set $\mathcal{M}_{\mathcal{A}}$ as $J(\mathcal{M}_{\mathcal{A}})$, ($[\boldsymbol{\Delta}]_{mm}=1 \iff m \in \mathcal{M}_{\mathcal{A}}$),
\begin{equation}
\label{eq:greedyobjective}
J(\mathcal{M}_{\mathcal{A}}) \triangleq f(\mathcal{M}_{\mathcal{A}}) - \alpha^{(t)}g(\mathcal{M}_{\mathcal{A}}).
\end{equation}
In each iteration, the algorithm first identifies a candidate set $\mathcal{C} \subseteq \mathcal{M}_{\mathcal{A}}$, which contains all active APs that can be deactivated without violating the QoS constraints for any UE. If this candidate set is non-empty, the algorithm then finds the best AP to sleep, $m^*$, by selecting the one whose removal maximizes the objective function $J(\cdot)$, as given by
\begin{equation}
\label{eq:greedyargmax}
m^* \leftarrow \arg\max_{m \in \mathcal{C}} J(\mathcal{M}_{\mathcal{A}} \setminus {m}).
\end{equation}
If this selection results in an improved objective value (i.e., $J(\mathcal{M}_{\mathcal{A}} \setminus \{m^*\}) > J(\mathcal{M}_{\mathcal{A}})$), the AP $m^*$ is deactivated by removing it from the active set $\mathcal{M}_{\mathcal{A}}$. The process repeats until no such improvement can be found.

\subsection{Whale Optimization Algorithm (WOA) for Subproblem 3}
\begin{table*}[!t]
\centering
\caption{Complexity Comparison per AO Iteration}
\label{tab:complexity_comparison}
\begin{tabular}{|c|c|c|}
\hline
\textbf{Component} & \textbf{Near-optimal solution (BnB + SCA + GP)} & \textbf{Low-complexity solution (Greedy + SCA + WOA)} \\ \hline
\begin{tabular}[c]{@{}l@{}}AP Selection\end{tabular} & 
  $O(2^M C_{\text{node}})$ & 
  $O(M^2  C_{\text{obj}})$ \\ \hline
\begin{tabular}[c]{@{}l@{}}Power Allocation\end{tabular} & 
  $O(I_{\text{SCA}}  (K^2 + C_{\text{LP}}(K, M_{\mathcal{A}})))$ & 
  $O(I_{\text{SCA}} (K^2 + C_{\text{LP}}(K, M_{\mathcal{A}})))$ \\ \hline
\begin{tabular}[c]{@{}l@{}}RIS Phase Shift \\ Optimization\end{tabular} & 
  $O(I_{\text{GP}} LN  K  (M_{\mathcal{A}} + LN))$ & 
  $O(I_{\text{WOA}}  N_{\text{pop}}  K  (M_{\mathcal{A}} + LN))$ \\ \hline
\end{tabular}
\end{table*}
To further reduce the complexity of the RIS phase optimization in Subproblem 3, we adopt a meta-heuristic approach, the WOA~\cite{mirjalili2016woa}. Inspired by the bubble-net feeding strategy of humpback whales, WOA is suitable for high-dimensional, non-convex phase shift optimization due to its global search capability and simplicity.

In WOA, each candidate solution (a vector of RIS phase shifts $\boldsymbol{\theta} = [\theta_{1,1}, \dots, \theta_{L,N}]$) is treated as a ``whale'' in the search space. The algorithm's behavior is modeled by three main phases: encircling prey, bubble-net attacking (exploitation), and searching for prey (exploration). Let $\boldsymbol{\theta}^*(t)$ be the best solution found so far at iteration $t$, and $\boldsymbol{\theta}_i(t)$ be the position of the $i$-th whale.

Whales identify and encircle the prey (the current best solution). This is mathematically modeled as:
\begin{equation}
    \boldsymbol{\theta}_i(t+1) = \boldsymbol{\theta}^*(t) - \mathbf{a}_c \odot \mathbf{D},
\end{equation}
where $\odot$ denotes the Hadamard product, and $\mathbf{D} = |\mathbf{c}_c \odot \boldsymbol{\theta}^*(t) - \boldsymbol{\theta}_i(t)|$. 
The coefficient matrices $\mathbf{a}_c$ and $\mathbf{c}_c$ are calculated as $\mathbf{a}_c = 2\mathbf{a} \odot \mathbf{r}_1 - \mathbf{a}$ and $\mathbf{c}_c = 2 \mathbf{r}_2$, where $\mathbf{a}$ is a control parameter that is linearly decreased from 2 to 0 to transition from an initial exploration phase to a final exploitation phase. 
The terms $\mathbf{r}_1$ and $\mathbf{r}_2$ are random vectors whose elements are drawn independently from a uniform distribution over [0, 1].

Bubble-net attacking is used to model the attacking behavior. A probability of 0.5 is used to switch between two mechanisms:
\begin{itemize}
    \item \textit{Shrinking Encircling Mechanism:} This is achieved by decreasing the value of $\mathbf{a}$. When $|\mathbf{a}_c| < 1$, the whale is forced to move towards the current best solution by the encircling mechanism.
    \item \textit{Spiral Updating Mechanism:} This mimics the helix-shaped movement of whales, defined as:
    \begin{equation}
        \boldsymbol{\theta}_i(t+1) = \mathbf{D}' e^{bl} \cos(2\pi l) + \boldsymbol{\theta}^*(t),
    \end{equation}
    where $\mathbf{D}' = |\boldsymbol{\theta}^*(t) - \boldsymbol{\theta}_i(t)|$, $b$ is a constant defining the spiral shape, and $l$ is a random number in $[-1, 1]$.
\end{itemize}

When $|\mathbf{a}_c| \geq 1$, the whale is forced to search for prey randomly, enhancing global exploration. This is achieved by updating the whale's position with respect to a randomly chosen whale $\boldsymbol{\theta}_{\text{rand}}$ instead of the best one:
\begin{equation}
    \boldsymbol{\theta}_i(t+1) = \boldsymbol{\theta}_{\text{rand}}(t) - \mathbf{a}_c \odot \mathbf{D}.
\end{equation}

The unit-modulus constraint for the RIS, $|\phi_{l,n}|=1$, is naturally handled by optimizing the phase angles $\theta_{l,n} \in [0, 2\pi)$ and then setting $\phi_{l,n} = e^{j\theta_{l,n}}$. After each position update, the phase angles are mapped back to the $[0, 2\pi)$ interval using the modulo operator to handle the periodicity of the search space. The fitness of each whale is evaluated by the sum rate objective $f(\boldsymbol{\Phi})$ under fixed AP selection and power allocation. The algorithm terminates after a fixed number of iterations.

Compared to gradient-based methods, WOA avoids complex derivations and matrix operations, making it highly suitable for large-scale RIS optimization. Although it does not guarantee optimality, it offers a favorable tradeoff between performance and complexity.





\subsection{Complexity Analysis}
The computational complexity of the proposed EE maximization framework is primarily determined by the algorithms employed within its two-layer iterative structure, consisting of an outer Dinkelbach loop and an inner AO loop. We analyze the complexity per AO iteration for two distinct algorithmic approaches: the near-optimal solution and the low-complexity solution, summarized in Table~\ref{tab:complexity_comparison}.

The near-optimal solution focuses on achieving higher performance by employing more computationally intensive techniques. AP selection is performed using a BnB algorithm, which, in the worst-case scenario for $M$ APs, exhibits a complexity of $O(2^M C_{\text{node}})$, where $C_{\text{node}}$ represents the cost of evaluating a single node in the search tree. Power allocation is handled via the SCA method over $I_{\text{SCA}}$ iterations. Each SCA iteration involves solving a convex subproblem, with a complexity of $O(I_{\text{SCA}} (K^2 + C_{\text{LP}}(K, M_{\mathcal{A}})))$, where $C_{\text{LP}}$ is the cost of solving the linear program. For the optimization of RIS phase shift, the GP method with numerical differentiation requires $O(I_{\text{GP}} LN C_{f\Phi})$, where $I_{\text{GP}}$ is the number of iterations of GP and $C_{f\Phi} = K (M_{\mathcal{A}} + LN)$ is the cost of a single sum rate evaluation. Consequently, the overall complexity of the near-optimal solution is dominated by the exponential term of the BnB algorithm.

The low-complexity solution prioritizes lower computational overhead for large-scale systems. AP selection is performed using a greedy heuristic, which has a complexity of $O(M^2 C_{\text{obj}})$, where $C_{\text{obj}}$ is the cost of one Dinkelbach objective evaluation. The power allocation is also handled by the SCA method, sharing the same complexity as in a near-optimal solution. For RIS phase optimization, the WOA is adopted. Its complexity is $O(I_{\text{WOA}} N_{\text{pop}} C_{f\Phi})$, where $I_{\text{WOA}}$ is the number of iterations of WOA, $N_{\text{pop}}$ is the population size and $C_{f\Phi}$ is the cost of the fitness evaluation (i.e., the sum rate). As all components are polynomial in the system dimensions, the overall complexity of the low-complexity solution is polynomial. This makes the low-complexity solution significantly more scalable with the number of APs, $M$, compared to the near-optimal solution.

\section{Numerical Results}\label{sec5}
In this section, we present numerical results to evaluate the  EE performance of the considered multi-RIS-aided  CF-mMIMO system.

\begin{table}[!t]
    \centering
    \renewcommand{\arraystretch}{1.2}
    \caption{Simulation parameters}
    \label{tab:simulation_parameters}
    \begin{tabular}{l|c}
        \hline
        \textbf{Parameter (symbol)} & \textbf{Value} \\ 
        \hline
        Carrier frequency ($f_c$) & 1.9 GHz \\ 
        System bandwidth ($B$) & 20 MHz \\
        RIS elements ($N$) & 64 ($8 \times 8$) \\
        Fixed AP power (\( P^{\text{AP,Fix}}_m \)) & 5 W \\ 
        Maximum radiated AP power (\( P_{\text{max}} \)) & 0.2 W \\ 
        Uplink transmission power ($p_\mathrm{u}$) & 0.1W\\
        CPU power consumption (\( P_{\text{CPU}}^{\text{Fix}}, P_{\text{CPU}}^{\text{Pre}} \)) & 5 W, 0.1 W/Gbps \\ 
        Fronthaul power consumption (\( P_{m}^{\text{FH,Fix}}, P_{m}^{\text{FH,Var}} \)) & 0.825 W, 0.01 W \\ 
        Power saving factor of dormancy $\varpi$ & 0.7 \\ 
        AP / UE antenna height & 12.5 m / 1.5 m \\ 
        RIS height & 13.5 m \\ 
        Max iterations & 30 \\
        SCA convergence tolerance ($\epsilon_{\text{SCA}}$) & $10^{-4}$ \\
        Gradient projection tolerance ($\varepsilon$) & $10^{-4}$ \\
        WOA population size ($N_{\text{pop}}$) & 30 \\ 
        \hline
    \end{tabular}
\end{table}
\subsection{Simulation Setup}\label{simulationsetup}
The simulation setup is based on the multi-RIS-aided CF-mMIMO system model from~\cite{channelmodelal2024performance} and the power consumption model from~\cite{sleepmodepowermodel}, with specific parameter values detailed in Table~\ref{tab:simulation_parameters}.
Each coherence interval comprises of $\tau_c = 1000$ symbols, resource block bandwidth of $200$ kHz, and coherence time of $1$ ms. The training time is $\tau_t = 20$ symbols and
$\tau_d = \tau_c - \tau_t$ is for downlink data transmission.
The number of APs $M$, UEs $K$, and RISs $L$ vary across simulation scenarios and are specified in the corresponding figure captions.
Each RIS has a static power of 0.064 W~\cite{pei2021ris}, and the minimum per-UE SE requirement is 1 bit/s/Hz.

To reflect a practical scenarios, the channel model incorporates probabilistic LoS conditions, where the small-scale fading is modeled as Rician when a LoS path exists and defaults to Rayleigh fading otherwise.
The existence of a LoS path in the AP-UE and RIS-UE links is determined by a two-stage process. First, a major blockage event is modeled by a Bernoulli random variable $\mathcal{B} \in \{0,1\}$, with a blockage ($\mathcal{B} = 0$) probability $p = 0.5$. Second, if no major blockage is present ($\mathcal{B} = 1$), a LoS path exists with a distance-dependent probability, as defined in~\cite{channelmodelal2024performance},
\begin{equation}
\Pr(\text{LoS} \mid \mathcal{B} = 1) = 
\begin{cases}
\frac{300 - d}{300}, & 0 < d < 300\,\text{m}, \\
0, & d > 300\,\text{m}.
\end{cases}
\end{equation}
The Rician $K$-factor, $\kappa_{ru,lk}$ and $\kappa_{au,mk}$, are set accordingly. If a LoS path exists, its value is calculated based on the distance $d$ as $K(\text{dB}) = 13 - 0.03d$~\cite{ozdogan2019massive}. In all NLoS cases (i.e., if $\mathcal{B}=0$ or $\Pr(\text{LoS} \mid \mathcal{B} = 1) = 0$), the $K$-factor is set to zero for NLOS case, which reduces the channel model to Rayleigh fading.


\subsection{Simulation Results}

\subsubsection{Convergence and Performance Comparison}
\begin{figure}[!t]
	\centering
	\includegraphics[width=\columnwidth]{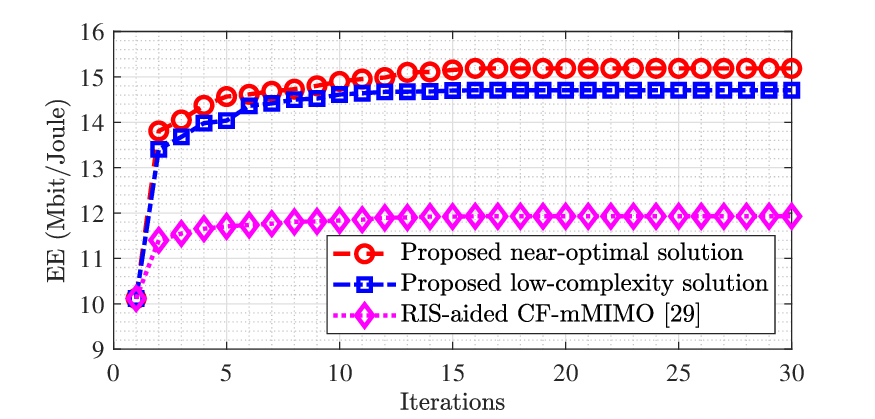}
	\caption{Convergence comparison of the proposed near-optimal solution, low-complexity solution, and the RIS-aided baseline, with $M = 10$, $K = 5$, and $L = 25$ in a $0.5\times0.5$ km$^2$ area.}
	\label{Iterations}
\end{figure}

To evaluate the effectiveness of our proposed joint optimization framework, we compare its convergence behavior against that of the RIS-aided optimization baseline~\cite{channelmodelal2024performance} in Fig.~\ref{Iterations}. The baseline in~\cite{channelmodelal2024performance} optimized RIS phase shifts for sum-rate maximization but does not consider an AP sleep-mode strategy.
The overall EE for all three schemes, which is updated in each Dinkelbach iteration, exhibits monotonic non-decreasing behavior. Our proposed algorithms converge to stable EE values within approximately 16 Dinkelbach iterations.
Both near-optimal and low-complexity solutions converge to substantially higher EE values than the RIS-aided baseline.
Furthermore, the figure illustrates the performance-complexity trade-off between the proposed schemes. The near-optimal solution achieves the highest EE, serving as our high-performance reference.
The low-complexity solution, while having lower computational complexity, achieves an EE that is
96.8\% of this reference's performance.
Because of this, all subsequent simulation results labeled ``Proposed scheme'' refer to the low-complexity solution.

\subsubsection{Impact of User Load}
\begin{figure}[!t]
	\centering
	\includegraphics[width=\columnwidth]{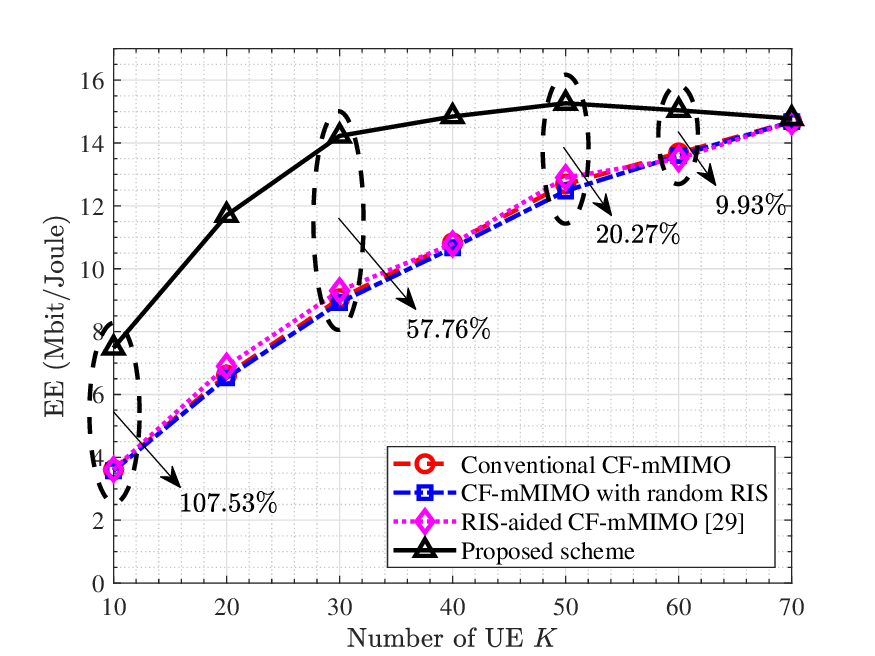}
	\caption{EE of an RIS-aided CF-mMIMO system and a CF-mMIMO system under varying number of UE $K$, with $M = 100$ and $L = 25$ in a $1\times1$ km$^2$ area.}
	\label{UEnumbA}
\end{figure}

To evaluate the impact of UEs load on system EE under dense AP deployment ($M = 100$), we compare the proposed scheme with the conventional CF-mMIMO (without RIS), CF-mMIMO with randomly configured RIS, and optimized RIS-aided CF-mMIMO~\cite{channelmodelal2024performance}. As shown in Fig.~\ref{UEnumbA}, EE increases with the number of UEs for all schemes, indicating that CF-mMIMO systems achieve higher EE under high UE load, while EE degrades significantly in low load scenarios.
This trend occurs because the sum rate grows substantially with more UEs due to multi-user diversity, while the total power consumption increases at a much slower rate. The large, fixed static power costs are amortized over a much higher total data throughput, making the system significantly more efficient at higher loads.

It is also observed that the proposed scheme consistently outperforms all baselines, particularly under low and moderate UE densities, with EE gains of 107.53\%, 57.76\%, 20.27\%, and 9.93\% over CF-mMIMO at $K = 10$, $K = 30$, $K = 50$, and $K = 60$, respectively.
This is primarily because none of these baselines are explicitly designed to maximize EE. They assume all APs are always active, resulting in substantial static power consumption that is particularly detrimental to EE in low and moderate load scenarios.
As $K$ increases, the performance gaps between our proposed scheme and the baselines become narrow since more APs need to remain active, limiting the benefits of sleep-mode optimization.
Notably, conventional CF-mMIMO outperforms the random RIS scheme, as the latter fails to provide constructive reflections and introduces additional power consumption.
Meanwhile, the method~\cite{channelmodelal2024performance} yields limited gains under the dense AP deployment scenario, where strong direct AP-UE links diminish the effectiveness of RIS-only optimization.

\subsubsection{Impact of AP Density}
\begin{figure}[!t]
    \centering
      \subfigure[]{
        \includegraphics[width=\columnwidth]{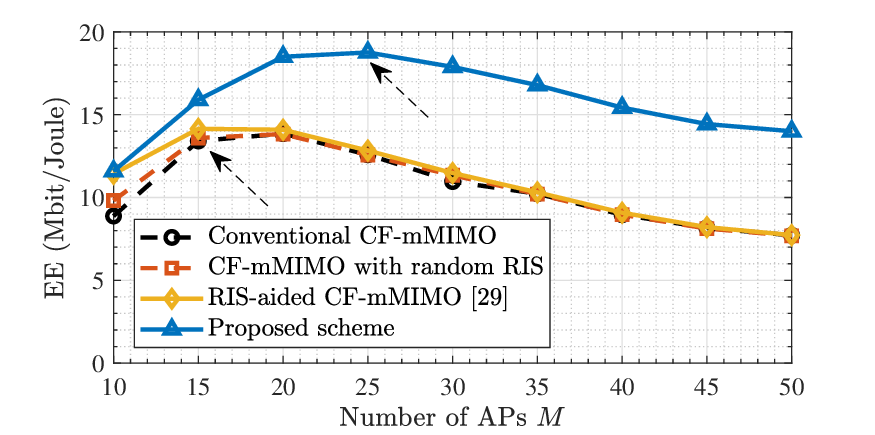}
        \label{APnumbA0.5}
    }  
        \subfigure[]{
        \includegraphics[width=\columnwidth]{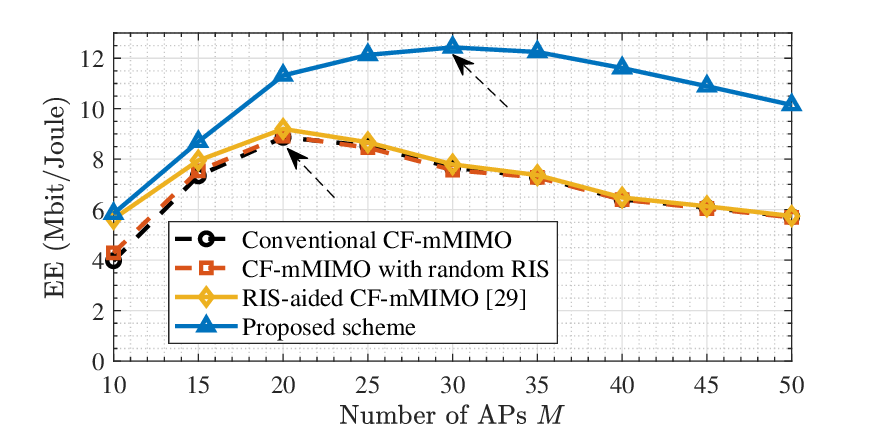}
        \label{APnumbA1}
    }
    \caption{EE of an RIS-aided CF-mMIMO system and a CF-mMIMO system under varying number of AP $M$, with $K = 10$ and $L = 5$.
(a) Scenario with $D = 0.5$ km.
(b) Scenario with $D = 1$ km.}
    \label{APnumbA}
\end{figure}
To evaluate the impact of AP deployment density on EE, we compare the proposed scheme with the baseline methods under varying numbers of APs in two different deployment area sizes, $D = 0.5$ km and $D = 1$ km, as illustrated in Fig.~\ref{APnumbA}. It is observed that for all schemes, the EE initially increases with $M$, reaches a peak, and then declines as the increasing static power consumption from a large number of APs outweighs the marginal rate improvements.


The proposed scheme consistently achieves the highest EE. Its advantage becomes more pronounced as $M$ increases, because the sleep-mode strategy allows the system to leverage the high spatial diversity of a dense AP deployment together with RIS without incurring the full energy cost. In contrast, schemes without sleep-mode suffer from a sharp EE degradation at high $M$, as their total power consumption grows linearly with the number of active APs.



Comparing Fig. \ref{APnumbA1} ($D = 1$ km) and Fig. \ref{APnumbA0.5} ($D = 0.5$ km), the optimal number of APs $M^*$ that maximize EE shifts to a higher value for all schemes considered in the larger deployment area. For instance, the peak EE value for the proposed scheme shifts from $M^* \approx 25$ in the compact area with $D = 0.5$ km to $M^* \approx 30$ in the larger area with $D = 1$ km, and similar rightward shifts are observed for the baseline curves. 
This is because in a smaller deployment area, a higher AP density provides users with significantly better direct AP–UE links, which rapidly diminishes the incremental rate benefit of adding another AP. Consequently, the point at which the energy cost of an additional AP outweighs its incremental rate benefit is reached sooner, leading to a lower optimal value of $M^*$.

On another note, the proposed scheme performs comparably to the RIS-aided scheme~\cite{channelmodelal2024performance} when $M = 10$. However, as the number of APs increases, the performance gap between the two schemes widens, with the proposed approach achieving significantly higher EE. This is because the proposed scheme leverages sleep mode, allowing redundant APs to be deactivated when not needed. In dense deployments, this flexibility enables better energy utilization without sacrificing coverage. The baseline, in contrast, lacks this adaptability, causing its energy consumption to increase linearly with $M$ while its performance gains saturate, ultimately leading to lower EE.


\subsubsection{Impact of the Number of RISs}
\begin{figure}[!t]
    \centering
    \subfigure[]{
        \includegraphics[width=\columnwidth]{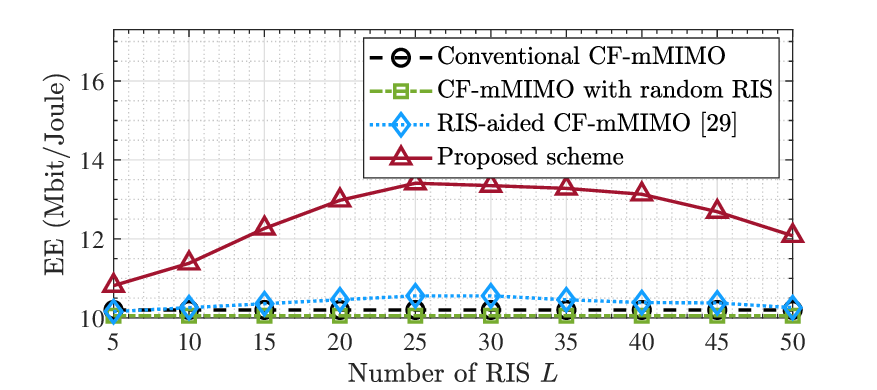}
        \label{fig:sub1}
    }
    
    \vspace{-0.35cm} 
    
    \subfigure[]{
        \includegraphics[width=\columnwidth]{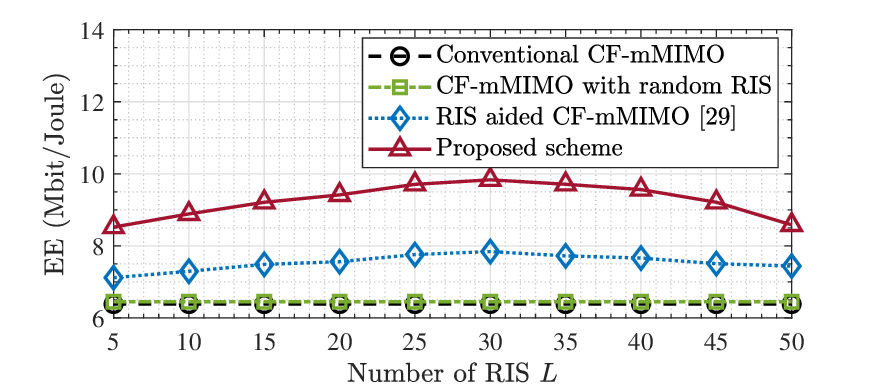}
        \label{fig:sub2}
    }
    
    \caption{EE of an RIS-aided CF-mMIMO system and a CF-mMIMO system under varying number of RISs $L$ in a $1\times1$ km$^2$ area.
(a) Scenario with $M = 40$ APs and $K = 20$ UEs. 
(b) Scenario with $M = 10$ APs and $K = 5$ UEs.}

    \label{fig:total}
\end{figure}

Finally, we evaluate the impact of the number of RISs $L$ on system EE. To investigate this impact under different network densities, we present two scenarios in Fig.~\ref{fig:sub1} a moderate density scenario ($M = 40$, $K = 20$), and Fig.~\ref{fig:sub2} a low density scenario ($M = 10$, $K = 5$).
It is noted that these two scenarios maintain the same ratio of APs to UEs ($M/K = 2$), which implies that the average AP resource level per UE is consistent across these two density regimes.
In both scenarios, the EE exhibits a similar trend with respect to $L$. EE initially increases, reaches a peak around $L = 30$, and subsequently declines. 
This behavior is attributed to the increased interference caused by dense RIS deployment and, albeit small, the increase in total RIS static power consumption. This highlights the need to match RIS deployment to network scale rather than simply maximizing their number.

Comparing the two scenarios, we also observe that in the denser network deployment, the relative gain of the proposed scheme over the scheme in~\cite{channelmodelal2024performance} that only deploys and optimizes RISs is more pronounced compared to the sparse deployment case. This is because the proposed scheme can flexibly activate necessary APs and put others to sleep, balancing performance and energy use. In contrast, the RIS-only scheme lacks such adaptability and incurs static overhead, which further validates the effectiveness of the proposed approach in future CF-mMIMO systems with dense infrastructure.

\subsubsection{Ablation Study of the Proposed Schemes}
\begin{figure}[!t]
    \centering
    \subfigure[]{
        \includegraphics[width=\columnwidth]{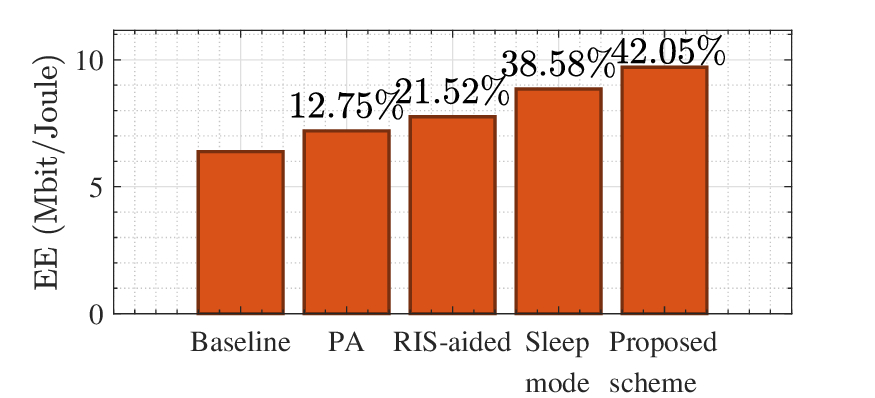}
        \label{Histogram1}
    }
\subfigure[]{
        \includegraphics[width=\columnwidth]{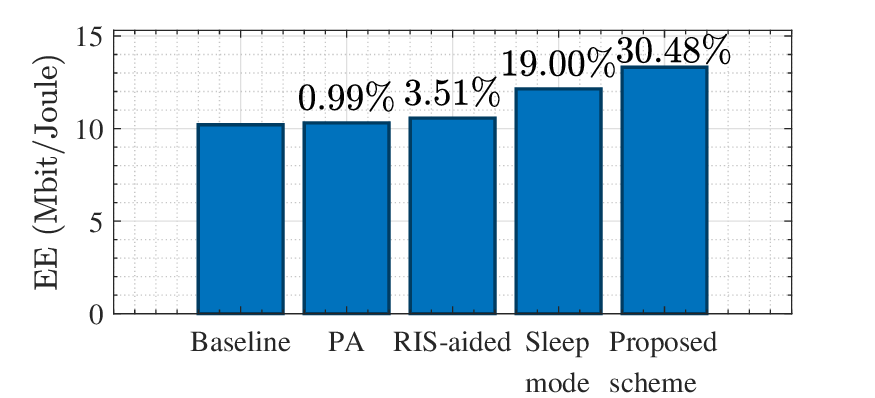}
        \label{Histogram2}
    }
    \caption{Ablation study on the EE contribution of different optimization components under two distinct system configurations with $L = 25$ in a $1\times1$ km$^2$ area.
(a) Sparse scenario ($M=10$, $K=5$). 
(b) Moderately dense scenario ($M=40$, $K=20$).}

    \label{Histogram}
\end{figure}
To quantify the impact of each element within our framework, we conduct an ablation study, with the results presented in Fig.~\ref{Histogram}, where the percentages are the improvements over the baseline. The performance of the proposed scheme is compared against a baseline (conventional CF-mMIMO without RIS) and three partially optimized schemes.
To ensure a fair and controlled comparison, the non-optimized variables in each partial scheme are set to a fixed, neutral configuration.
When optimizing power allocation (PA) only, all APs are assumed to be active, and the RIS phase shifts are set to random values. When optimizing RIS phase shifts only, all APs remain active, and a fixed, heuristic-based power allocation scheme is employed throughout the process. When optimizing sleep-mode only, the RIS phase shifts are set to random values, and the same fixed power allocation scheme is used.
This analysis uses two different network densities to validate the robustness.
Each optimization component provides a noticeable EE gain, with the sleep-mode strategy yielding the most substantial improvement among the single-optimization schemes. This confirms that reducing static power consumption via AP deactivation is the most effective means to enhance EE.

Furthermore, a comparison between Fig.~\ref{Histogram1} and Fig.~\ref{Histogram2} reveals a key insight into the role of RIS. In the sparse scenario, the introduction of RIS optimization provides a EE gain of approximately 21.52\%. However, in the moderately dense scenario, the same RIS configuration yields a smaller gain of 3.51\%. This is because in denser networks with a high number of active APs, the abundance of high-quality direct AP-UE links diminish the relative contribution of the weaker, two-hop RIS-reflected paths.

\section{Conclusion}\label{sec6}
This paper presents a novel energy-efficient framework for multi-RIS-aided CF-mMIMO networks by co-optimizing AP sleep mode, power allocation and RIS passive beamforming. 
The proposed solution, tackled via a Dinkelbach and AO scheme, intelligently deactivates underutilized APs while leveraging RISs to maintain coverage, thereby drastically improving EE under low-to-moderate UE loads.
Two algorithms were developed: a BnB-based solution targeting near-optimal performance, suitable for smaller systems or offline benchmarks, and a scalable greedy-based heuristic, favored for larger networks and real-time applications due to its lower complexity.
Simulation results show that the low-complexity solution can achieve $96.8\%$ of the performance of the near-optimal solution, offering an excellent trade-off between performance and computational cost.
Overall, the proposed scheme improves EE for low and moderate UE scenarios compared to benchmark, thereby providing a powerful and practical solution for enhancing the energy efficiency of future CF-mMIMO aided wireless networks.


\appendices

\section{}
\label{appendix:proof_of_rate_theorem}

In this appendix, we provide a detailed derivation for the interference power term resulting from imperfect CSI. This term appears in the denominator of the UE $k$'s SINR expression, as shown in~(\ref{eq:sinr_theorem}).

The interference arises because the ZF precoder is designed based on the estimated channel $\mathbf{\hat{G}}$, while the actual transmission occurs over the true channel $\mathbf{g}_k$. The resulting interference signal received at UE $k$ can be denoted by $I_k$:
\begin{equation}
    I_k \triangleq \check{\mathbf{g}}_k^T \Delta \mathbf{\hat{G}}^{*} \left( \mathbf{\hat{G}}^T \Delta \mathbf{\hat{G}}^{*} \right)^{-1} \mathbf{P} \mathbf{s},
    \label{eq:interference_signal}
\end{equation}
where $\check{\mathbf{g}}_k = \mathbf{g}_k - \hat{\mathbf{g}}_k$ is the channel estimation error vector. Our objective is to compute the average power of this interference, $\mathbb{E}\{|I_k|^2\}$.

By leveraging the assumption that the data symbols $\mathbf{s}$ are normalized and uncorrelated (i.e., $\mathbb{E}\{\mathbf{s}\mathbf{s}^H\} = \mathbf{I}_K$), the expression for the interference power becomes:
\begin{equation}
\begin{aligned}
    \mathbb{E}\{|I_k|^2\} = &\mathbb{E} \left\{\left|\check{\mathbf{g}}_k^{T} \Delta \mathbf{\hat{G}}_k^{*} \left( \mathbf{\hat{G}}_k^{T} \Delta \mathbf{\hat{G}}_k^{*} \right)^{-1} \mathbf{P}\right|^2 \right\} \\
         = &\mathbb{E} \bigg\{\check{\mathbf{g}}_k^{T} \Delta \mathbf{\hat{G}}_k^{*} \left( \mathbf{\hat{G}}_k^{T} \Delta \mathbf{\hat{G}}_k^{*} \right)^{-1} \mathbf{P} \mathbf{P}^H \\ &\left( \left( \mathbf{\hat{G}}_k^{T} \Delta \mathbf{\hat{G}}_k^{*} \right)^{-1} \right)^H \mathbf{\hat{G}}_k^{T} \Delta \check{\mathbf{g}}_k^{*} \bigg\}.
\end{aligned}
\end{equation}

Since the term inside the expectation is a scalar, we can take its trace without changing its value. We then apply the cyclic property of the trace operator $\operatorname{tr}(\mathbf{\cdot})$ to rearrange the matrices for subsequent steps. Hence,
\begin{equation}
\begin{aligned}
    \mathbb{E}\{|I_k|^2\} = \mathbb{E}\bigg\{  \operatorname{tr}\bigg\{  &\mathbf{P} \mathbf{P}^H \left( \left( \mathbf{\hat{G}}_k^{T} \Delta \mathbf{\hat{G}}_k^{*} \right)^{-1} \right)^H  \mathbf{\hat{G}}_k^{T} \Delta \check{\mathbf{g}}_k^{*} \check{\mathbf{g}}_k^{T} \\ &\Delta \mathbf{\hat{G}}_k^{*} \left( \mathbf{\hat{G}}_k^{T} \Delta \mathbf{\hat{G}}_k^{*} \right)^{-1} \bigg\} \bigg\}.
\end{aligned}
\label{eq:trace_form}
\end{equation}

Due to the linearity of both the trace and expectation operators, we can swap their order. Since the estimation error $\check{\mathbf{g}}_k$ is uncorrelated with the estimate $\mathbf{\hat{G}}$, the expectation over the error term can be derived as follows:
\begin{equation}
    \begin{aligned}
    \mathbb{E}\{|I_k|^2\} = \operatorname{tr}\bigg\{   \mathbf{P} \mathbf{P}^H \mathbb{E}&\bigg\{ \left( \left( \mathbf{\hat{G}}_k^{T} \Delta \mathbf{\hat{G}}_k^{*} \right)^{-1} \right)^H \mathbf{\hat{G}}_k^{T} \Delta \check{\mathbf{g}}_k^{*} \check{\mathbf{g}}_k^{T} \\& \Delta \mathbf{\hat{G}}_k^{*} \left( \mathbf{\hat{G}}_k^{T} \Delta \mathbf{\hat{G}}_k^{*} \right)^{-1} \bigg\} \bigg\} \\
         = \operatorname{tr}\bigg\{   \mathbf{P} \mathbf{P}^H  \mathbb{E}&\bigg\{\left( \left( \mathbf{\hat{G}}_k^{T} \Delta \mathbf{\hat{G}}_k^{*} \right)^{-1} \right)^H \mathbf{\hat{G}}_k^{T} \Delta \mathbb{E}\left(\check{\mathbf{g}}_k^{*} \check{\mathbf{g}}_k^{T} \right) \\ &\Delta \mathbf{\hat{G}}_k^{*} \left( \mathbf{\hat{G}}_k^{T} \Delta \mathbf{\hat{G}}_k^{*} \right)^{-1} \bigg\} \bigg\},
         \end{aligned}
\end{equation}
where the inner expectation is, by definition, the covariance matrix of the channel estimation error, $\mathbf{C}_k = \mathbb{E}\left(\check{\mathbf{g}}_k^{*} \check{\mathbf{g}}_k^{T} \right)$, as defined in Section~\ref{ULtrainingchannelestimation}.

The remaining expectation term now corresponds precisely to the definition of the interference vector $\boldsymbol{\gamma}_k$ given in (17). We arrive at the final expression
\begin{equation}
         \mathbb{E}\{|I_k|^2\} =\sum^K_{i=1} \eta_i \gamma_{k,i}.
\end{equation}
This completes the derivation of the interference power term used in~\eqref{eq:sinr_theorem}.

\section{}\label{Derivation}
The rate function for UE $k$ is given by $R_k(\mathbf{P}) = B \log_2\left(1+ \frac{p_k}{ D_k(\mathbf{P})}\right)$, where $D_k(\mathbf{P}) = \sum_{i=1}^K p_i \gamma_{k,i}(\boldsymbol{\Phi}^\star) + \sigma_{d,k}^2$. This can be rewritten as a difference of two functions:
\begin{equation}
R_k(\mathbf{P}) = B \left[ \log_2\left(p_k + D_k(\mathbf{P})\right) - \log_2\left(D_k(\mathbf{P})\right) \right].
\end{equation}
Let $f_k(\mathbf{P}) = \log_2(p_k + D_k(\mathbf{P}))$ and $g_k(\mathbf{P}) = \log_2(D_k(\mathbf{P}))$. Since $p_k + D_k(\mathbf{P})$ and $D_k(\mathbf{P})$ are affine functions of $\mathbf{P}$, and $\log_2(\cdot)$ is a concave function, both $f_k(\mathbf{P})$ and $g_k(\mathbf{P})$ are concave functions of $\mathbf{P}$.

The expression $f_k(\mathbf{P}) - g_k(\mathbf{P})$ represents the difference of two concave functions, which is generally non-concave. To obtain a tractable convex approximation, we construct a lower bound for $R_k(\mathbf{P})$ by leveraging the properties of concave functions. A concave function is always upper-bounded by its first-order Taylor expansion. At a given point $\mathbf{P}^{(s)}$, we have:
\begin{equation}
g_k(\mathbf{P}) \leq g_k(\mathbf{P}^{(s)}) + \nabla_{\mathbf{P}} g_k(\mathbf{P}^{(s)})^T (\mathbf{P} - \mathbf{P}^{(s)}).
\end{equation}
Therefore, we can lower-bound $R_k(\mathbf{P})$ by replacing $g_k(\mathbf{P})$ with its affine upper bound:
\begin{equation}
\begin{aligned}
    R_k(\mathbf{P}) &= B[f_k(\mathbf{P}) - g_k(\mathbf{P})] \\&\geq B[f_k(\mathbf{P}) - (g_k(\mathbf{P}^{(s)}) +  \nabla_{\mathbf{P}} g_k(\mathbf{P}^{(s)})^T (\mathbf{P} - \mathbf{P}^{(s)}))].
\end{aligned}
\end{equation}
This lower bound is denoted as $\tilde{R}_k^{(s)}(\mathbf{P})$. To find its explicit form, we need the gradient of $g_k(\mathbf{P})$. 
The $j$-th element of the gradient vector $\nabla_{\mathbf{P}} g_k(\mathbf{P})$ is:

\begin{equation}
    \begin{aligned}
\frac{\partial g_k(\mathbf{P})}{\partial p_j} = &\frac{\partial}{\partial p_j} \log_2\left(\sum_{i=1}^K p_i \gamma_{k,i}(\boldsymbol{\Phi}^\star) + \sigma_{d,k}^2\right)\\
= &\frac{1}{\ln 2 \left(\sum_{i=1}^K p_i \gamma_{k,i}(\boldsymbol{\Phi}^\star) + \sigma_{d,k}^2\right)} \\ &\times \frac{\partial}{\partial p_j}\left(\sum_{i=1}^K p_i \gamma_{k,i}(\boldsymbol{\Phi}^\star)\right)\\
= &\frac{\gamma_{k,j}(\boldsymbol{\Phi}^\star)}{\ln 2 D_k(\mathbf{P})}.
\end{aligned}
\end{equation}

Evaluating the gradient at $\mathbf{P}^{(s)}$ and substituting it back, we obtain the affine lower bound $\tilde{R}_k^{(s)}(\mathbf{P})$ as presented in (\ref{eq:rate_lower_bound}). This bound is a sum of a concave function $B f_k(\mathbf{P})$ and an affine function, which preserves concavity. This ensures that the resulting SCA subproblem (\ref{eq:sca_power_subproblem}) is a convex optimization problem.

\bibliographystyle{IEEEtran}
\bibliography{IEEEabrv,EEsleep}

\end{document}